\documentclass[a4paper,11pt]{article}

\pdfoutput=1
\usepackage{amsmath}
\usepackage{graphicx}

\usepackage[english]{babel}
\usepackage[utf8]{inputenc}
\usepackage{pdflscape}
\usepackage{enumerate}
\usepackage{amsbsy}
\usepackage{amsmath} 
\usepackage{graphics}
\usepackage{mathrsfs}
\usepackage{wrapfig}
\usepackage{mathtools}

\usepackage{amsfonts}
\usepackage{pstricks}
\usepackage{color}
\usepackage{setspace}

\usepackage{jcappub_ver} 

\usepackage{makecell}

\newcommand{\bea}{\begin{eqnarray}} \newcommand{\eea}{\end{eqnarray}}
\newcommand{\el}{\nonumber \\}
\newcommand{\re}[1]{(\ref{#1})}

\newcommand{\pat}{\partial}

\renewcommand{\sec}[1]{section \ref{#1}}

\newcommand{\tab}[1]{table \ref{#1}}

\renewcommand{\a}{\alpha}
\renewcommand{\b}{\beta}
\renewcommand{\c}{\gamma}
\renewcommand{\d}{\delta}

\newcommand{\m}{\mu}

\newcommand{\ha}{\frac{1}{2}}

\newcommand{\rmd}{\mathrm{d}}

\newcommand{\ie}{i.e.\ }
\newcommand{\eg}{e.g.\ }

\newcommand{\half}{\ha}

\newcommand{\pd}{\partial}
\newcommand{\cd}{\nabla}

\newcommand{\lcd}{\mathring{\cd}}

\newcommand{\lcdq}{\overset{q}{\lcd}}
\newcommand{\sg }{\sqrt{-g}}

\newcommand{\lcR}{\mathring{R}}
\newcommand{\sq}{\sqrt{-q}}
\usepackage{accents}
\usepackage{tensor}
\newcommand{\TI}{\indices}

\newcommand{\cR}{\hat{R}} 
\DeclareRobustCommand{\dR}{\accentset{\Delta}{R}} 
\DeclareRobustCommand{\hR}{\accentset{\sim}{R}} 
\newcommand{\pR}{\accentset{P}{R}} 
\newcommand{\tf}{\d \tilde f}

\newcommand{\dL}{\d L}
\newcommand{\bL}{\bar{L}}

\newcommand{\dq}{\d q }
\newcommand{\bq}{\bar{q}}

\newcommand{\dS}{\d S}
\newcommand{\bS}{\bar{S}}

\newcommand{\dP}{\d P}

\newcommand{\bcd}{\bar{\cd}}

\newcommand{\dqt}{ h }
\newcommand{\dSt}{ s }

\title{Stability of non-degenerate Ricci-type Palatini theories}

\author{Jaakko Annala}
\author{and Syksy R\"{a}s\"{a}nen}

\affiliation{University of Helsinki, Department of Physics and Helsinki Institute of Physics \\ P.O. Box 64, FIN-00014 University of Helsinki, Finland}

\emailAdd{jaakko.annala@helsinki.fi}
\emailAdd{syksy.rasanen@iki.fi}

\abstract{We study the stability of theories where the gravitational action has arbitrary algebraic dependence on the three first traces of the Riemann tensor: the Ricci tensor, the co-Ricci tensor, and the homothetic curvature tensor. We collectively call them Ricci-type tensors. We allow arbitrary coupling to matter. We consider the case when the connection is unconstrained, and the cases when either torsion or non-metricity is assumed to vanish. We find which combinations of Ricci-type tensors lead to new degrees of freedom around Minkowski and FLRW space, and when there are ghosts. None of the theories with new degrees of freedom are healthy, except for two previously known cases where there is a single new vector. We find that projective invariance is not a sufficient condition for a theory to be ghost-free.}

\begin{document}

\begin{flushleft}
	\hfill		 HIP-2022-33/TH \\
\end{flushleft}
 
\setcounter{tocdepth}{3}

\setcounter{secnumdepth}{3}

\maketitle

\section{Introduction} \label{sec:intro}

Renormalisation in quantum field theory in curved spacetime requires higher order curvature terms, and they also appear in proposed ultraviolet completions of general relativity \cite{Birrell:1982ix, Zwiebach:1985uq, Zumino:1985dp, Deser:1986xr, Eichhorn:2020mte}. In the metric formulation of gravity, non-linear curvature terms generally lead to higher derivative equations of motion that have new degrees of freedom that suffer from the Ostrogradski instability \cite{Stelle:1977ry, Simon:1990ic, MuellerHoissen:1991, Nunez:2004ts, Chiba:2005nz, Woodard:2006nt}. (It is possible that this classical instability does not persist in quantum theory \cite{Donoghue:2021eto}.) In 4 spacetime dimensions, non-linear curvature terms always (boundary terms aside) lead to higher derivative equations of motion and new degrees of freedom, but not all of them are unstable. In 4 dimensions, the most general healthy action with no matter and no dependence on derivatives of the Riemann tensor depends on the curvature only via Ricci scalar. With arbitrary algebraic dependence on the Ricci scalar, the action has one extra non-ghost scalar degree of freedom \cite{Sotiriou:2008rp}. When matter is added, there are further stable cases, such as the Horndeski and beyond Horndeski theories that include scalar fields \cite{Kobayashi:2019hrl}.

In the Palatini formulation, the metric and the connection are independent degrees of freedom \cite{Einstein:1925, Hehl:1976, Hehl:1978, Hehl:1981}. The equations of motion are therefore second order for any action that is algebraic in the Riemann tensor and where the connection appears only via the Riemann tensor.\footnote{Fermions, unlike scalar fields and gauge fields, necessarily couple to a connection in the action, although this can be taken to be the Levi--Civita connection rather than the full connection \cite{Hehl:1976, Randono:2005, Freidel:2005sn, Perez:2005pm, Mercuri:2006um, Mercuri:2006wb, Bojowald:2007nu, Kazmierczak:2008, Shaposhnikov:2020gts, Shaposhnikov:2020aen}.} For this reason higher order curvature terms do not necessarily lead to new degrees of freedom, and new degrees of freedom do not involve the Ostrogradski instability. However, the theory can still be unstable because kinetic terms or gradient terms can have the wrong sign, or potentials can be tachyonic or unbounded from below \cite{Vasilev:2017zyc,BeltranJimenez:2019acz, BeltranJimenez:2020sqf, Percacci:2020ddy, Jimenez-Cano:2022sds}. (Fields with a kinetic term of the wrong sign may not be a problem in the quantum theory \cite{Donoghue:2019ecz, Donoghue:2020mdd}.) As in the metric case, requiring the theory to be stable greatly restricts the action. However, unlike in the metric case, the most general stable theory is not known even when there is no matter. In addition to the Riemann curvature tensor, the geometrical properties of the manifold are described by the non-metricity tensor and the torsion tensor. There are hence many more gravitational terms than in the metric case, and terms that individually correspond to an unstable theory can lead to a stable theory when combined \cite{Vasilev:2017zyc,Percacci:2020ddy,Jimenez-Cano:2022sds}.

Many Palatini theories have been considered in the literature. It is known that a theory with arbitrary algebraic dependence on the symmetric part of the Ricci tensor does not contain any new degrees of freedom \cite{Afonso:2017, Galtsov:2018xuc, Annala:2021zdt, Olmo:2022rhf}. If the torsion is taken to be zero, a theory with arbitrary algebraic dependence on the antisymmetric part of the Ricci tensor is equivalent to normal gravity with an additional healthy vector degree of freedom \cite{Vitagliano:2010pq, Olmo:2013lta, BeltranJimenez:2019acz}. The stability properties of some parity-invariant theories quadratic in the curvature, non-metricity, and torsion and are also known \cite{Alvarez:2017spt, Alvarez:2018lrg, Percacci:2020ddy, Marzo:2021esg}.

In the Palatini formulation, the Einstein--Hilbert action is invariant under the projective transformation of the connection \cite{Hehl:1978}. It has been found that also in some extended actions requiring projective invariance guarantees stability \cite{Aoki:2019rvi, BeltranJimenez:2019acz, BeltranJimenez:2020sqf}. Stable Palatini theories without projective symmetry are known, so it is not a necessary condition for stability \cite{Rasanen:2018b, BeltranJimenez:2019acz, BeltranJimenez:2020sqf}. Furthermore, it is expected that in general projective symmetry is not enough to guarantee stability, see \eg chapter 8 of \cite{Delhom:2021bvq}.

We extend previous work by considering the stability of theories where the connection enters only via the first contractions of the Riemann tensor. There are three independent first contractions: the Ricci tensor, the co-Ricci tensor, and the homothetic curvature tensor. We collectively call them Ricci-type curvature tensors, or just Ricci-type tensors.

Hamiltonian analysis is the definitive way to establish the dynamical content and stability of a theory \cite{Belenchia:2016bvb}. However, with these complicated theories the analysis can be quite involved. We take a simpler route. We perform a Legendre transformation and introduce auxiliary fields to make the theory linear in the Riemann tensor. Expanding around Minkowski space or the Friedmann--Lema\^{\i}tre--Robertson--Walker (FLRW) universe to first order, we solve for the connection and insert the solution back into the action. Having reduced the theory to metric gravity with the Einstein--Hilbert action plus minimally coupled matter, we then look at the kinetic sector of the new matter fields.

This method has its limits. We are restricted to non-degenerate theories, meaning theories for which the Legendre transformation is invertible. We also cannot find the dynamical content of the theory in general, only around specific backgrounds. If the theory is found to contain unstable degrees of freedom, this is sufficient to rule it out, but the reverse does not hold. It is possible for a theory to be well-behaved around one background but unstable around another. A theory may also contain new degrees of freedom around one background but not another. This can indicate a strong coupling problem \cite{Hindawi:1995cu, Golovnev:2018wbh, Pookkillath:2020iqq, BeltranJimenez:2020lee}. We find examples of both situations. We also find a projectively symmetric theory where perturbations around Minkowski space are unstable, showing that projective symmetry is not a sufficient condition for stability.

In section \ref{sec:dof} we first introduce the relevant geometrical quantities and the action. We then make the Legendre transformation and introduce the auxiliary fields. We look at the degrees of freedom and stability in three cases: the general case when the connection is unconstrained, the zero torsion case, and the zero non-metricity case. We consider perturbations around the Minkowski background, and in the general case also around the spatially flat FLRW background, and find when kinetic terms have the wrong sign. We summarise our findings and outline open questions in section \refeq{sec:conc}. Some technical details are relegated to appendices \ref{app:con} and \ref{app:general}.

\section{Degrees of freedom}\label{sec:dof}

\subsection{Geometrical quantities and the action}\label{sec:geom}

\subsubsection{Torsion and non-metricity}

We work in the Palatini formulation, where the metric and the connection $\Gamma^\c_{\a\b}$ are independent degrees of freedom. The connection is related to the covariant derivative as $\cd_\b A^\a=\pd_\b A^\a + \Gamma^\a_{\b\c} A^\c$, $\cd_\b A_\a=\pd_\b A_\a - \Gamma^\c_{\b\a} A_\c$. We can decompose the connection as
\begin{equation} \label{Con_decomp}
	\Gamma^\c_{\a\b} = \overset{g}{\mathring\Gamma}{}^\c_{\a\b} + L\TI{^\c_{\a\b}} \ ,
\end{equation}
where $\overset{g}{\mathring\Gamma}{}^\c_{\a\b}$ is the Levi--Civita connection of the metric $g_{\a\b}$, and $L\TI{^\c_{\a\b}}$ is the related distortion tensor. In this decomposition we can define the Levi--Civita connection with respect to any metric, it does not have to be the physical spacetime metric. Both the Levi--Civita connection and the distortion tensor depend on the choice of the metric, while their sum does not.

The non-metricity tensor and the two non-metricity vectors defined with the metric $g_{\a\b}$ are
\bea
	\label{def_Q} \overset{g}{Q}_{\c\a\b} \equiv \cd_\c g_{\a\b} = - 2 L_{(\a|\c|\b)} \ , \quad \overset{g}{Q}_\c \equiv \overset{g}{Q}_{\c\a\b}  g^{\a\b} \ , \quad \overset{g}{\hat{Q}}_\b \equiv \overset{g}{Q}_{\c\a\b} g^{\c\a} \ ,
\eea
where $L_{\c\a\b}\equiv g_{\c\mu} L^\mu{}_{\a\b}$. The torsion tensor and the torsion vector are defined independently of the metric as
\bea 
	\label{def_T} T\TI{^\c_{\a\b}}\equiv 2\Gamma^\c_{[\a\b]} = 2 L\TI{^\c_{[\a\b]}} \ , \quad T_\b \equiv T\TI{^\a_{\b\a}} \ .
\eea
The distortion can be written in terms of non-metricity and torsion as
\begin{equation}
	L\TI{^\c_{\a\b}} = J\TI{^\c_{\a\b}} + K\TI{^\c_{\a\b}} \ ,
\end{equation}
where the disformation and the contortion are, respectively,
\begin{equation}\label{JK}
	J^\a{}_{\b\c} \equiv \half g^{\a\mu} ( \overset{g}{Q}_{\mu\b\c}  - \overset{g}{Q}_{\c\mu\b} - \overset{g}{Q}_{\b\mu\c} ) \ , \quad K\TI{^\a_{\b\c}} \equiv \half ( T\TI{^\a_{\b\c}} + g_{\c\mu} g^{\a\nu} T^\mu{}_{\nu\b} + g_{\b\mu} g^{\a\nu} T^\mu{}_{\nu\c} ) \ .
\end{equation}
Note that $J_{\a\b\c}=J_{\a(\b\c)}$ and $K\TI{^{\a}_\b^{\c}}=K\TI{^{[\a}_\b^{\c]}}$, with indices lowered with $g_{\a\b}$ and raised with its inverse. The tensors $L^\a{}_{\b\c}$, $J^\a{}_{\b\c}$, and $K\TI{^\a_{\b\c}}$ all depend on the choice of the metric, but to avoid overburdening the notation we do not label them with $^g$, as it is usually clear from the context which metric is used. Transformations of the metric shift pieces of the connection between $\overset{g}{\mathring\Gamma}{}^\c_{\a\b}$, $J^\a{}_{\b\c}$, and $K\TI{^\a_{\b\c}}$.

\subsubsection{The Ricci-type tensors}

The Riemann tensor depends only on the connection,
\begin{equation}\label{Rie_def}
	R^{\a}{}_{\b\c\d} \equiv \pd_\c \Gamma^{\a}_{\d\b}-\pd_\d\Gamma^{\a}_{\c\b} + \Gamma^{\a}_{\c\mu} \Gamma^{\mu}_{\d\b} - \Gamma^{\a}_{\d\mu}\Gamma^{\mu}_{\c\b} \ .
\end{equation}
With the connection decomposition \eqref{Con_decomp}, we can write \re{Rie_def} in terms of Levi--Civita and distortion contributions as
\bea \label{Rie_dec}
  R^{\a}{}_{\b\c\d} = \overset{g}{\mathring R}{}^{\a}{}_{\b\c\d} + 2 \overset{g}{\mathring \nabla}_{[\c} L^\a{}_{\d]\b} + 2 L^\a{}_{[\c|\mu|} L^\mu{}_{\d]\b} \ ,
\eea
where $\overset{g}{\mathring{}}$ denotes a quantity defined with the Levi--Civita connection $\overset{g}{\mathring\Gamma}{}^\c_{\a\b}$. 

For a general connection the only symmetry of the Riemann tensor is antisymmetry in the last two indices. There are thus three independent first contractions of the Riemann tensor: the Ricci tensor, the co-Ricci tensor, and the homothetic curvature tensor, respectively defined as
\begin{align} \label{Riccitensors}
	R_{\a\b} &\equiv R\TI{^\mu_{\a\mu\b}} \el
	\cR\TI{^\a_\b} &\equiv g^{\mu\nu}R\TI{^\a_{\mu\nu\b}} \el
	\hR_{\a\b} &\equiv R\TI{^\mu_{\mu\a\b}} \ .
\end{align}
We collectively refer to these three quantities (and their linear combinations) as Ricci-type curvature tensors, or just Ricci-type tensors. The homothetic curvature tensor $\cR\TI{^\a_\b}$ depends on the metric, while the other two Ricci-type tensors depend only on the connection. The full contraction of the Riemann tensor is unique, since $R\equiv g^{\a\b}R_{\a\b}=-\cR\TI{^\a_\a}$ and $g^{\a\b}\hR_{\a\b}=0$. 

Inserting the decomposition of the connection in terms of non-metricity and torsion \re{Con_decomp}--\re{JK} into the decomposition \re{Rie_dec} of the Riemann tensor, we can write the co-Ricci tensor as
\begin{equation} \label{ricci_diff}
	\cR_{\a\b} = - R_{\a\b} + 2 g^{\mu\nu}\nabla_{[\b} \overset{g}{Q}_{\mu]\nu\a} - T^{\mu\nu}{}_\b \overset{g}{Q}_{\mu\nu\a} \ .
\end{equation}
Instead of the co-Ricci tensor, it will turn out to be more convenient to use the average of the co-Ricci tensor and the Ricci tensor,
\begin{equation} \label{ricci_av}
	\dR_{\a\b} \equiv \half ( \cR_{\a\b}+R_{\a\b} ) = g^{\mu\nu}\nabla_{[\b} \overset{g}{Q}_{\mu]\nu\a} - \ha T^{\mu\nu}{}_\b \overset{g}{Q}_{\mu\nu\a} \ .
\end{equation}
Note that $g^{\a\b}\dR_{\a\b}=0$. Similar decomposition for the homothetic curvature tensor gives
\bea \label{hRdec}
  \hR_{\a\b} &=& \pat_{[\b}\overset{g}{Q}_{\a]} \ .
\eea
So the homothetic curvature tensor is the exterior derivative of the non-metricity vector. This decomposition obscures the fact that $\hR_{\a\b}$ is independent of the metric, as is transparent from the definition \re{Riccitensors}. We can choose to do the decomposition \re{hRdec} with respect to any metric. So a sufficient (but not necessary) condition for $\hR_{\a\b}$ to vanish is that the manifold is such that there exists any symmetric tensor $q_{\a\b}$ that is non-degenerate ($\det q_{\a\b} \neq 0$) and covariantly constant ($\nabla_\c q_{\a\b}=0$).

\subsubsection{Projective transformation}

An important property of the Einstein--Hilbert action in the Palatini formulation is invariance under the projective transformation \cite{Hehl:1978}
\begin{equation}\label{projective}
	\Gamma^\c_{\a\b} \to \Gamma^\c_{\a\b} + \delta\TI{^\c_\b} V_\a \ ,
\end{equation}
where $V_\a$ is an arbitrary vector. Under this transformation non-metricity and torsion change as
\begin{align}
		\overset{g}{Q}_{\c\a\b} &\to \overset{g}{Q}_{\c\a\b} - 2 V_{\c}g_{\a\b} \el
		T\TI{^\c_{\a\b}} &\to T\TI{^\c_{\a\b}} + 2\delta\TI{^\c_{[\b}}V_{\a]} \ . 
\end{align}
Thus the non-metricity and torsion vectors change as 
\bea \label{vectorpro}
		\overset{g}{Q}_{\a} &\to& \overset{g}{Q}_{\a} - 8V_\a \el
		\overset{g}{\hat{Q}}_{\a} &\to& \overset{g}{\hat{Q}}_{\a} - 2V_\a \el
		T_\a &\to& T_\a + 3 V_\a \ .
\eea
The traceless part of the non-metricity tensor and the part of the torsion tensor that does not contribute to $T_\a$ are invariant. Therefore, if the theory is invariant under the projective transformation, we can completely exchange torsion for non-metricity if the torsion tensor can be written in terms of the torsion vector, $T\TI{^\c_{\a\b}} = - \frac{2}{3} \delta\TI{^\c_{[\a}}T_{\b]}$. Equivalently, if non-metricity is involved only via the non-metricity vectors $\overset{g} Q_\a$ and $\overset{g}{\hat{Q}}$, and they are related by $\overset{g} Q_\a=4 \overset{g}{\hat{Q}}$, it can be exchanged for torsion.\footnote{This is the case in $f(R)$ gravity \cite{Iosifidis:2018zjj}.}

The Riemann tensor transforms as
\bea \label{Riemannpro}
  R^\a{}_{\b\c\d}\to R^\a{}_{\b\c\d}+ 2 \d^\a{}_\b \partial_{[\c} V_{\d]} \ ,
\eea
so the Ricci-type tensors \re{Riccitensors} and \re{ricci_av} transform as
\begin{align} \label{Riccipro}
	& R_{\a\b}\to R_{\a\b} + 2\pd_{[\a}V_{\b]} \el
	& \cR_{\a\b} \to \cR_{\a\b} + 2\pd_{[\a}V_{\b]} \el
	& \hR_{\a\b}\to \hR_{\a\b} + 8\pd_{[\a}V_{\b]} \el
	& \dR_{\a\b}\to\dR_{\a\b}+2\pd_{[\a}V_{\b]} \ .
\end{align}
The symmetric parts of the Ricci-type tensors are invariant under the projective transformation. Also, the antisymmetric combination
\bea \label{projsym_R}
  \pR_{\a\b} \equiv (\alpha+4\beta)R_{[\a\b]} - \alpha \dR_{[\a\b]} - \beta \hR_{\a\b} \ ,
\eea
is projectively invariant; here $\alpha$ and $\beta$ are any projectively invariant quantities.

\subsubsection{Action and Legendre transformation}\label{sec:act}

We consider the most general Palatini theory where the action depends algebraically on the Ricci-type tensors, and has no other dependence on the connection. We can take the action to depend on any set of independent combinations of the Ricci-type tensors. For convenience we choose $R_{\a\b},\dR_{(\a\b)},\pR_{\a\b}$, and $\hR_{\a\b}$. The action reads
\begin{equation} \label{action1}
	S = \int \rmd^4 x \sg \half F(g_{\a\b},R_{\a\b},\dR_{(\a\b)},\pR_{\a\b},\hR_{\a\b},\Psi,\pd\Psi) \ ,
\end{equation}
where $g\equiv\det{g_{\a\b}}$, and $\Psi$ and $\pd\Psi$ collectively denote matter fields and their derivatives (which can be higher than first order), with arbitrary coupling to the Ricci-type tensors. In order to have all of the degrees of freedom of the Ricci-type tensors represented, we must have $\a\neq0$ in $\pR_{\a\b}$ defined in \re{projsym_R}. Also, the terms in $\pR_{\a\b}$ proportional to $\b$ are already separately variables in the action. So without loss of generality we set $\a=1$, $\b=0$. We perform a Legendre transformation to make the action linear in the Ricci-type tensors \cite{Magnano:1987zz, Ferraris:1988zz, Magnano:1990qu, Koga:1998un, Afonso:2017}
\bea \label{actionL}
		S &=& \int \rmd^4 x \sg \half \left[ F\left(g_{\a\b},\Sigma_{\a\b},\Delta_{\a\b},\Pi_{\a\b},\Theta_{\a\b},\Psi,\pd\Psi\right) + \frac{\pd F}{\pd \Sigma_{\a\b}} ( R_{\a\b} -\Sigma_{\a\b} ) \right. \el
		&& \left. + \frac{\pd F}{\pd \Delta_{\a\b}} ( \dR_{(\a\b)} -\Delta_{\a\b} )+ \frac{\pd F}{\pd \Pi_{\a\b}} ( \pR_{\a\b} -\Pi_{\a\b} ) + \frac{\pd F}{\pd \Theta_{\a\b}}( \hR_{\a\b} -\Theta_{\a\b} )  \right] \ ,
\eea
where $\Sigma_{\a\b}$, $\Delta_{\a\b}=\Delta_{(\a\b)}$, $\Pi_{\a\b}=\Pi_{[\a\b]}$, and $\Theta_{\a\b}=\Theta_{[\a\b]}$ are auxiliary fields. Let us check when the action \re{actionL} is equivalent to the original action \re{action1}. Let us first consider $\Sigma_{\a\b}$. If the action is linear in $\Sigma_{\a\b}$, equivalence follows trivially. Let us then assume that this is not the case, so $\pat^2 F/(\pat\Sigma_{\a\b}\pd\Sigma_{\c\d})\neq0$. Variation with respect to $\Sigma_{\a\b}$ gives
\bea\label{var_Sigma}
	\frac{\pd^2 F}{\pd \Sigma_{\a\b}\pd\Sigma_{\c\d}}\left( R_{\c\d} - \Sigma_{\c\d} \right) = 0 \ .
\eea
 If $\pat^2 F/(\pat\Sigma_{\a\b}\pd\Sigma_{\c\d})$ is invertible with respect to the indices $\a\b$ (or equivalently $\c\d$), \re{var_Sigma} gives $\Sigma_{\a\b}=R_{\a\b}$. The invertibility requirement imposes a constraint on the form of $F$. We need the same assumption also for the second partial derivatives of $F$ with respect to $\Delta_{\a\b}$, $\Pi_{\a\b}$, and $\Theta_{\a\b}$. We call theories where $F$ satisfies these constraints non-degenerate. For such theories we also get $\Delta_{\a\b}=\dR_{(\a\b)}$, $\Pi_{\a\b} = \pR_{\a\b}$, and $\Theta_{\a\b}=\hR_{\a\b}$. Inserting these solutions back into \re{actionL} recovers the action \re{action1}.

We now introduce the field redefinitions (we use units where the Planck mass is unity)
\bea
	\label{redefs} \sg \frac{\pd F}{\pd \Sigma_{(\a\b)}} &\equiv& \sq q^{\a\b} \ , \quad \sg \frac{\pd F}{\pd \Sigma_{[\a\b]}} \equiv \sq B^{\a\b} \\
	\label{redefS} \sg \frac{\pd F}{\pd \Delta_{\a\b}} &\equiv& \sq S^{\a\b} \ , \quad \sg \frac{\pd F}{\pd \Pi_{\a\b}} \equiv \sq P^{\a\b} \\
	\label{redefH}  \sg \frac{\pd F}{\pd \Theta_{\a\b}} &\equiv& \sq H^{\a\b} \ ,
\eea
where $q\equiv1/\det q^{\a\b}=\det q_{\a\b}$, with $q_{\a\b}$ being the inverse of $q^{\a\b}$. (The inverse has to exist for the field redefinition \re{redefs} to be consistent.) In terms of these new fields, the action \re{actionL} reads
\bea \label{general_legendre}
  S &=& \int \rmd^4 x \sq \half \Bigg[ q^{\a\b}R_{(\a\b)} + B^{\a\b}R_{[\a\b]} + S^{\a\b} \dR_{(\a\b)} + \Pi^{\a\b}\pR_{\a\b} + H^{\a\b}\hR_{\a\b} \el \label{general_legendre_f}
  &&+ \frac{\sg}{\sq} F - (q^{\a\b}+B^{\a\b})\Sigma_{\a\b} - S^{\a\b} \Delta_{\a\b} - P^{\a\b}\Pi_{\a\b} - H^{\a\b}\Theta_{\a\b} \Bigg] \el
  &\equiv& \int \rmd^4 x \sq \half \Bigg[ R^\a{}_{\b\c\d} f_\a{}^{\b\c\d} \el
  && + \frac{\sg}{\sq} F - (q^{\a\b}+B^{\a\b})\Sigma_{\a\b} - S^{\a\b} \Delta_{\a\b} - P^{\a\b}\Pi_{\a\b} - H^{\a\b}\Theta_{\a\b} \Bigg] \ ,
\eea
where we have used the definitions \re{Riccitensors} and \re{ricci_av} of the Ricci-type tensors to gather the factors multiplying the Riemann tensor into the term
\bea\label{def_gen_f}
  f_\a{}^{\b\c\d} &\equiv& \left( q^{\b[\d} + B^{\b[\d} + \ha S^{\b[\d} +\ha P^{\b[\d} \right) \d^{\c]}{}_\a \el
  && + \ha S^{\mu[\d}  g^{\c]\b} g_{\mu\a} - \frac{1}{2} P^{\mu[\d}  g^{\c]\b} g_{\mu\a} + H^{\c\d} \d^\b{}_\a \ .
\eea
The field $q_{\a\b}$ plays the role of the metric, and we have the symmetric field $S^{\a\b}$, and the antisymmetric fields $ B^{\a\b}$, $P^{\a\b}$, and $H^{\a\b}$. The old fields $g_{\a\b}, \Sigma_{\a\b},\Delta_{\a\b},\Pi_{\a\b}$, and $\Theta_{\a\b}$ are solved algebraically in terms of the new fields $q_{\a\b},B^{\a\b},S^{\a\b},P^{\a\b}$, $H^{\a\b}$, and the matter fields $\Psi$ (and their derivatives) from the field redefinitions \re{redefs}--\re{redefH}.

The action is linear in the Riemann tensor, which is the only place where the connection appears. So the action is quadratic in the connection. Variation with respect to the connection hence gives a linear equation of motion, which we can solve, and insert the solution back into the action. The kinetic terms for the remaining fields then indicate the number of degrees of freedom and whether they are ghosts or healthy. We look at the degrees of freedom in three cases: when the connection is unconstrained, when torsion is assumed to vanish, and when non-metricity is assumed to vanish.

\subsection{General case}

\subsubsection{Equation of motion}

We start with the case when no constraints are imposed on the connection. Projective transformation of the new Legendre transformed action then plays an important role. Projective invariance of the new action \re{general_legendre} is distinct from possible projective invariance of the original action \re{action1}. Whether or not the original formulation of the theory is projectively invariant depends on which combinations of Ricci-type tensors the action contains. In the new action, part of the physics related to the connection in the original action has been transferred to the auxiliary fields, which are taken to be projectively invariant. Under a projective transformation, the Riemann tensor transforms as \re{Riemannpro}, so the new  action \re{general_legendre} changes as 
\bea \label{actionpro}
	\delta_{P}S &=& \int \rmd^4 x \sq f_\a{}^{\a\c\d}\pd_\c V_\d \el
	&=& - \int \rmd^4 x \sq V_\d \overset{q}{\lcd}{}_\c f_\a{}^{\a\c\d} \ ,
\eea
where on the last line we have switched $\pd_\a V_\b$ to $\overset{q}{\lcd}_\a V_\b$ using the fact that the contribution of the Levi--Civita connection vanishes due to the antisymmetry $f_\a{}^{\b\c\d}=f_\a{}^{\b[\c\d]}$, done a partial integration, and dropped a boundary term. The Legendre transformed action is thus projectively invariant if
\bea \label{procon}
	0 &=& \overset{q}{\lcd}_\c f_\a{}^{\a\c\d} = \overset{q}{\lcd}_\c ( B^{\c\d} + 4H^{\c\d} ) \ .
\eea
The field $S^{\a\b}$ does not appear in the constraint, corresponding to the fact that only the antisymmetric parts of the Ricci-type tensors change under the projective transformation \re{Riccipro}. 

Variation of the new action \re{general_legendre} with respect to the connection gives
\bea \label{variation}
  \d S &=& \int \rmd^4 x \sq \half \d R^\a{}_{\b\c\d} f_\a{}^{\b\c\d} \el
  &=& \int \rmd^4 x \sq \left( \nabla_\c \d\Gamma^\a_{\d\b} + \ha T^\mu{}_{\c\d} \d\Gamma^\a_{\mu\b} \right) f_\a{}^{\b\c\d} \el
  &=& \int \rmd^4 x \sq \left[ - \left( \nabla_\c + \ha \overset{q}{Q}_\c + T_\c \right) \d^\d{}_\mu + \ha T^\d{}_{\c\mu} \right] f_\a{}^{\b\c\mu} \d\Gamma^\a_{\d\b} \ ,
\eea
where we have used the definition \re{Rie_def} of the Riemann tensor, applied the identity $\nabla_\a\sq=\ha\sq Q_\a$, and done a partial integration. Putting the variation to zero gives the connection equation of motion
\bea \label{gen_con_eom}
  0 &=& \left[ ( \nabla_\c + \ha \overset{q}{Q}_\c + T_\c ) \d^\d{}_\mu - \ha T^\d{}_{\c\mu} \right] f_\a{}^{\b\c\mu} \el
  &=& \lcdq_\c f_\a{}^{\b\c\d} - L^\m{}_{\c\a}f_\m{}^{\b\c\d}+L^\b{}_{\c\m}f_\a{}^{\m\c\d} \ .
\eea
The $\d^\a{}_\b$ trace of this equation gives the constraint \re{procon}. So the equation of motion imposes a constraint on the matter fields that makes the action projectively invariant. Correspondingly, the projective symmetry leaves undetermined one linear combination of the traces of the distortion $L\TI{^\a_{\b\c}}$, corresponding to a linear combination of $\overset{q}{Q}_\a$, ${\overset{q}{\hat Q}}_\a$, and $T_\a$.

As equation \re{gen_con_eom} is linear, it is possible to solve $L\TI{^\a_{\b\c}}$ exactly. However, the general solution is rather messy. We consider perturbations around a background, and solve the equation of motion to first order.

\subsubsection{Minkowski space} \label{sec:Min}

We expand the original metric as $g_{\a\b} = \eta_{\a\b} + \delta g_{\a\b}$, where $\eta_{\a\b}$ is the Minkowski metric and $\delta g_{\a\b}$ is a perturbation. The new metric is similarly expanded as $q_{\a\b} = \eta_{\a\b} + \delta q_{\a\b}$. The connection $\Gamma^\c_{\a\b}$ and the fields $B^{\a\b},S^{\a\b},P^{\a\b}$, and $H^{\a\b}$ are first order small. To first order in perturbations, $f_\a{}^{\b\c\d}$ defined in \eqref{def_gen_f} reads
\begin{equation}
	f_\a{}^{\b\c\d} =  q^{\b[\d} \d^{\c]}{}_\a + \tf_\a{}^{\b\c\d} \ ,
\end{equation}
where
\bea\label{def_gen_f_1st}
  \!\!\!\!\!\!\!\!\!\!\!\!  \tf_\a{}^{\b\c\d} &\equiv& \left( B^{\b[\d} + \ha S^{\b[\d} +\ha P^{\b[\d} \right) \d^{\c]}{}_\a+ \ha S^{\mu[\d}  \eta^{\c]\b} \eta_{\mu\a} - \frac{1}{2} P^{\mu[\d}  \eta^{\c]\b} \eta_{\mu\a} + H^{\c\d} \d^\b{}_\a \ .
\eea
To first order, the connection equation of motion \eqref{gen_con_eom} simplifies to
\begin{equation}\label{general_eom_1st}
	- \overset{q}{Q}_{\c\a\b} + \eta_{\a\c} \overset{q}{\hat Q}_\b+ \eta_{\a[\b} ( \overset{q}{Q}_{\c]} + 2 T_{\c]}  ) + T_{\a\b\c} = 2\pd_\d\tf_{\c\b\a}{}^\d \ ,
\end{equation}
where indices are raised and lowered with the Minkowski metric. Solving \re{general_eom_1st} and choosing the torsion vector $T_\a$ as the part of the connection left undetermined due to projective invariance, we get
\begin{eqnarray}
	    \label{solQandT}
		\overset{q}{Q}_{\c\a\b} &=& \eta_{\a\b}\left( -\frac{2}{3}T_\c+\pd_\d\tf_{\c\m}{}^{\m\d} + \frac{1}{3}\pd_\d\tf_{\m\c}{}^{\m\d}\right) + 2\pd_\d \tf_{(\a\b)\c}{}^\d - 2\pd_\d\tf_{(\a|\c|\b)}{}^\d - 2\pd_\d\tf_{\c(\a\b)}{}^\d \el
		T_{\c\a\b} &=& \frac{2}{3}\eta_{\c[\b}T_{\a]} - 2\pd_\d \tf_{[\a\b]\c}{}^\d + 2\pd_\d\tf_{[\a|\c|\b]}{}^\d+ 2\pd_\d\tf_{\c[\a\b]}{}^\d + \frac{4}{3}\eta_{\c[\a}\pd^\m\tf^\d{}_{\b]\d\m} \ .
\end{eqnarray}
Inserting the solution \eqref{solQandT} into the action \re{general_legendre} and expanding to second order we obtain the action
\bea \label{action_2nd_gen}
		S &=& \int \rmd^4 x \half \Bigg[ \lcR + \half \pd_\c \tf_\a{}^{\b\a\c} \pd^\m \tf^\d{}_{\b\d\m} +\pd_\c\tf^{\a\b\c\d}\pd^\m\tf_{\a\b\d\m} \el
		&& -2\pd_\c\tf^{\a\b\c\d}\pd^\m\tf_{\b\d\a\m} + \half \pd_\c\tf^{\a\b}{}_\b{}^\c \pd^\m\tf_\a{}^\d{}_{\d\m} -\pd_\c\tf_\b{}^{\a\b\c}\pd_\m\tf_{\a\d}{}^{\d\m} \el
    && + \frac{2}{3}T_\a \pd_\b \tf_\c{}^{\c\a\b} + \pd_\b \tf_\a{}^{\a\b\c}\Big( \pd_\m \tf_{\c\d}{}^{\d\m} + \frac{1}{3}\pd_\m \tf_\d{}_\c{}^{\d\m} + 3\pd^\m \tf^\d{}_{\d\c\m} \Big) \el
		&& + \frac{\sg}{\sq} F -  (q^{\a\b}+B^{\a\b})\Sigma_{\a\b} - S^{\a\b} \Delta_{\a\b} - P^{\a\b}\Pi_{\a\b} - H^{\a\b}\Theta_{\a\b} \Bigg] \ .
\eea
Matter fields end up being minimally coupled because we use $\dR_{\a\b} = \half ( \cR_{\a\b}+R_{\a\b} )$ rather than $\cR_{\a\b}$ as a variable. Plugging in \eqref{def_gen_f_1st}, we get
\bea \label{action_2nd_gen_fields}
  S &=& \int \rmd^4 x \half \Bigg[ \lcR  - \frac{7}{2}\pd_\a {B}^{\a\b}\pd^\c {B}_{\c\b} -\frac{1}{4}\pd_\c {B}_{\a\b}\pd^\c {B}^{\a\b}  - \frac{166}{3}\pd_\a {H}^{\a\b}\pd^\c {H}_{\c\b} -\frac{82}{3} \pd_\a {B}^{\a\b}\pd^\c {H}_{\c\b} \el
  &&  + \frac{1}{3} ( - 2 T_\b + \pd^\c S_{\c\b} + \pd^\c P_{\c\b} ) ( \pd_\a B^{\a\b} + 4\pd_\a H^{\a\b} ) \el
  && + \frac{\sg}{\sq} F -  (q^{\a\b}+B^{\a\b})\Sigma_{\a\b}  - S^{\a\b} \Delta_{\a\b} - P^{\a\b}\Pi_{\a\b} - H^{\a\b}\Theta_{\a\b} \Bigg] \ .
\eea
The kinetic terms of ${B}^{\a\b}$ (corresponding to $R_{[\a\b]}$) give rise to a pseudovector degree of freedom that is unstable regardless of the presence of other fields; see appendix \ref{app:general} for details. In particular, if the action were to depend on the projectively invariant combination $4R_{[\a\b]}-\hR_{\a\b}$, we would also obtain a pseudovector ghost, as $\hR_{\a\b}$ contributes only to the vector sector of the action and thus cannot cancel the instability. So projective invariance does not guarantee stability.  An action that depends only on $\dR_{[\a\b]}$ also leads to an unstable theory in the same way as $R_{[\a\b]}$. But since both $R_{[\a\b]}$ and $\dR_{[\a\b]}$ contribute to the pseudovector sector their contributions can cancel each other. Indeed, the tensor $\pR_{\a\b} = R_{[\a\b]} - \dR_{[\a\b]}$ does not lead to a kinetic term (there is no kinetic term for $P^{\a\b}$), so the theory may be stable if the action contains $R_{[\a\b]}$ only in this combination.

If the action \eqref{action_2nd_gen_fields} does not depend on $B^{\a\b}$, its stability depends on the form of $F$, due to the term coming from the projective mode of the connection that has the form of a Lagrange multiplier.\footnote{In earlier versions of the paper, we incorrectly used the constraint \re{procon} in the action. The $T_\b$ term in the action now acts like a Lagrange multiplier. (It is the projective mode, and does not appear in the cases when we apply constraints to the connection, as they break the projective symmetry.) So the equation of motion of $T_\b$ enforces the constraint, but $T_\b$ also appears in the equation of motion of the fields $B^{\a\b}$ and $H^{\a\b}$, and can lead to new degrees of freedom. Because $B^{\a\b}$ and $H^{\a\b}$ can depend on the derivative of $T_\b$, the $F$ term can lead to kinetic terms, not just mass terms.} The action either involves no new degrees of freedom or contains one propagating vector field, which can be stable or unstable depending on the form of $F$; see appendix \ref{app:general} for details. The propagating vector originating in the homothetic curvature tensor $\hR_{\a\b}$ is known in the literature \cite{PhysRevD.56.7769}.

We see that the kinetic terms corresponding to different Ricci-type tensors can cancel each other, so it is not possible to conclude that one of them cannot be present in a stable theory without considering the others. As we do not get kinetic terms for $S^{\a\b}$ nor $P^{\a\b}$, we cannot draw conclusions about the stability of an action that contains the corresponding tensors $\dR_{(\a\b)}$ or $\pR_{\a\b}$. However, we next show that for $\dR_{(\a\b)}$ this feature is peculiar to Minkowski space, as around a cosmological background it leads to new degrees of freedom, which may imply a strong coupling problem \cite{Hindawi:1995cu, Golovnev:2018wbh, Pookkillath:2020iqq, BeltranJimenez:2020lee}.

\subsubsection{FLRW universe} \label{sec:FLRW}

We look at the kinetic terms of $S^{\a\b}$ and $P^{\a\b}$ (corresponding to $\dR_{\a\b}$ and $\pR_{\a\b}$ respectively) around a cosmological background. It is cumbersome to expand the action around a FLRW background in the case when we keep dependence on all Ricci-type tensors. We therefore consider a simplified case, where only $g^{\a\b} R_{\a\b}$, $\dR_{(\a\b)}$, and $\pR_{\a\b}$ appear,
\bea\label{action_FLRW}
	S = \int \rmd^4 x \sg \half F(g_{\a\b},g^{\a\b}R_{\a\b},\dR_{(\a\b)},\pR_{\a\b},\Psi,\pd\Psi) \ .
\eea
With this simplification the metrics $g_{\a\b}$ and $q_{\a\b}$ are conformal to each other, which makes expansion around a FLRW background more tractable. Were we to allow $F$ to depend on $R_{(\a\b)}$, the metrics would be related by a disformal transformation \cite{Afonso:2017, Galtsov:2018xuc, Annala:2021zdt}. Performing the Legendre transformation and introducing field redefinitions as before, the auxiliary fields are defined in \eqref{redefS}, \eqref{redefH}, and 
\bea
	\sg \frac{\pd F}{\pd \Sigma}g^{\a\b} &\equiv& \sq q^{\a\b} \ ,
\eea
where $\Sigma$ is the auxiliary field corresponding to $g^{\a\b} R_{\a\b}$. Now the action becomes
\bea \label{action_FLRW_2}
  \!\!\!\!\!\!\!\! S &=&  \int \rmd^4 x \sq \half \Bigg[ R^\a{}_{\b\c\d} f_\a{}^{\b\c\d} + \frac{\sg}{\sq} F -  \Sigma - S^{\a\b} \Delta_{\a\b} -P^{\a\b}\Pi_{\a\b}  \Bigg] \ ,
\eea
with
\bea \label{f_FLRW}
  \!\!\!\!\!\!\!\! f_\a{}^{\b\c\d} &=& \left( q^{\b[\d} + \ha S^{\b[\d}  +\frac{1}{2}P^{\b[\d} \right) \d^{\c]}{}_\a + \ha S^{\mu[\d}  g^{\c]\b} g_{\mu\a} - \frac{1}{2} P^{\mu[\d}  g^{\c]\b} g_{\mu\a} \el
  &=& \left( q^{\b[\d} + \ha S^{\b[\d} +\frac{1}{2}P^{\b[\d}\right) \d^{\c]}{}_\a + \ha S^{\mu[\d}  q^{\c]\b} q_{\mu\a} - \frac{1}{2} P^{\mu[\d}  q^{\c]\b} q_{\mu\a}  \ ,
\eea
where on the second line we have taken into account that $g_{\a\b}$ is conformal to $q_{\a\b}$. Because of this, the gravity sector of the action, where the kinetic terms can arise, does not explicitly depend on $g_{\a\b}$.

We now expand the action \eqref{action_FLRW_2} around a spatially flat FLRW background. The fields are split into background plus perturbations as
\bea
q_{\a\b} &=& \bq_{\a\b} + \dq_{\a\b} \ , \qquad\ \ \bq_{\a\b} = a(\eta)^2\eta_{\a\b} \el
	S^{\a\b} &=& \bS^{\a\b} + \dS^{\a\b} \ , \qquad L^{\c}{}_{\a\b} = \bL^{\c}{}_{\a\b}+\dL^{\c}{}_{\a\b}\ ,
\eea
where overbar denotes a background quantity, and $\eta$ is conformal time. We decompose $\bS^{\a\b}$ as
\bea \label{anz_bS}
	\bS^{\a\b} = S_1(\eta) \bq^{\a\b} + S_2(\eta) u^\a u^\b \ ,
\eea
where $u^{\a}=(a^{-1},0,0,0)$. As the field $P^{\a\b}$ is antisymmetric, it is zero on the background, so $P^{\a\b}=\dP^{\a\b}$. When we substitute \eqref{anz_bS} into the action \re{action_FLRW_2}, the action turns out not to depend on $S_1$.\footnote{Therefore, for a maximally symmetric background, where there is no preferred time direction and hence no $u^\a$, we will not get a kinetic term for $S^{\a\b}$, same as in Minkowski space.}

We decompose the background of the distortion tensor as
\bea\label{anz_bL}
\bL^{\c}{}_{\a\b} = L_1(\eta) u^{\c}\bq_{\a\b}+L_2(\eta) u_{\a}\delta^{\c}{}_{\b}+L_3(\eta) u_{\b}\delta^{\c}{}_{\a}+L_4(\eta) u^{\c}u_{\a}u_{\b} + L_5(\eta) \epsilon^{\c}{}_{\a\b\d} u^{\d} \ ,
\eea
where $\epsilon_{\a\b\c\d}$ is the Levi--Civita tensor. We substitute this into the action \re{action_FLRW_2}, vary the action with respect to the functions $L_i$, and solve the resulting equations to find the solution $\bL^\c{}_{\a\b} = \bL^\c{}_{\a\b}(a,S_2)$. The details are given in appendix \ref{app:con}. Inserting the background solution back into the action \re{action_FLRW_2}, the Riemann tensor term becomes
\bea\label{bg_act}
  S &\supset& \int \rmd^4 x a^4 \left[ \half {\lcR}(\bq) + \frac{3}{4} H^2S_2^2 + \frac{3S_2^2}{16 a^2(S_2^2 - 4)}S_2^\prime S_2^\prime + \frac{3 HS_2}{4 a}S_2^\prime \right] \ ,
\eea
where ${}^\prime \equiv \rmd / \rmd\eta$, $H\equiv a^\prime/a^2$, and ${\lcR}(\bq)$ is the Levi--Civita Ricci scalar of the background metric $\bar q_{\a\b}$. The kinetic term has the wrong sign when $|S_2|<2$, so the value $S_2=0$ is unstable, and the field runs to $S_2=\pm2$. This is transparent if we write the action in terms of a scalar field with a canonical kinetic term, $\chi=\sqrt{3|S_2^2-4|/8}$.

Let us now consider perturbations. We have to solve the connection to first order from \eqref{gen_con_eom}. We sketch here the main points of the calculation and summarise the results, with details given in appendix \ref{app:con}. The equation of motion for the connection to first order in perturbations has the form
\bea\label{con1st}
  \dL^\sigma{}_{\mu\nu} M_\sigma{}^{\mu\nu}{}_\a{}^{\b\c}(a,S_2) + \rho_{\a}{}^{\b\c}(a,S_2,\dq_{\mu\nu},\dS^{\mu\nu},\dP^{\mu\nu}) = 0 \ ,
\eea
where $M_\sigma{}^{\mu\nu}{}_\a{}^{\b\c}$ depends only on the background quantities, while the source term $\rho_{\a}{}^{\b\c}$ also depends on the perturbations and their derivatives. Solving the connection is straightforward but tedious. Substituting the solution back into the action again gives the kinetic terms. The kinetic part of the action expanded to second order has the form
\bea \label{action_kinetic_FLRW}
  S &\supset& \int \rmd^4 x a^4 \Big( K_{(1)}^{\mu\a\b\nu\c\d} \pd_\mu\dq_{\a\b}\pd_\nu\dq_{\c\d} + K_{(2)}^{\mu\a\b\nu\c\d}\pd_\mu\dS_{\a\b}\pd_\nu\dS_{\c\d}   + K_{(3)}^{\mu\a\b\nu\c\d}\pd_\mu\dq_{\a\b}\pd_\nu\dS_{\c\d} \Big) \ , \el
\eea
where $K_{(i)}^{\mu\a\b\nu\c\d}$ depend on the background quantities. The field $P^{\a\b}$ does not get kinetic terms.

To determine whether or not there are ghosts we have to diagonalise the kinetic terms. To facilitate this, we decompose the fields $\dq_{\a\b}$ and $\dS^{\a\b}$ into scalar, vector, and tensor modes. To second order, these decouple from each other, and we can look at each of them individually. If any are unstable, the theory is unhealthy. The tensor modes are the simplest: we have $\dq_{ij} = \dqt_{ij}$, with $\dqt{}^i{}_i = 0, \ \pd^i\dqt{}_{ij} = 0$, and $\dS^{ij} = \dSt^{ij}$, with $\dSt{}^i{}_i = 0, \ \pd_i\dSt{}^{ij} = 0$. Applying this decomposition in \eqref{action_kinetic_FLRW}, we get the tensor sector
\bea
  S &\supset& \int \rmd^4 x a^4 \left[ \frac{ S_2^2 + 4 }{16a^2} \pd_0 \dqt_{ij}\pd_0\dqt^{ij} -  \frac{1}{4}\pd_k\dqt_{ij}\pd^k \dqt^{ij} \right.  \left.+ \frac{S_2^2}{4a^2(S_2^2 - 4)}\pd_0\dSt_{ij}\pd_0\dSt^{ij}-\frac{S_2}{4a^2} \pd_0 \dqt_{ij}\pd_0\dSt^{ij} \right] \ , \el
\eea
where the spatial indices are lowered with $\bar q_{ij}$ and raised with its inverse. Diagonalising, we obtain
\bea
 S &\supset& \int \rmd^4 x a^4 \left[ \frac{1}{2a^2}\pd_0\dqt_{ij}\pd_0\dqt^{ij} - \frac{1}{4}\pd_k\dqt_{ij}\pd^k \dqt^{ij} + \frac{S_2^2}{4a^2(S_2^2 - 4)} \pd_0\dSt_{ij}\pd_0\dSt^{ij} \right] \ .
\eea
The factor in front of the kinetic term of $\dSt^{ij}$ has the same form as the background solution, and has the wrong sign when $|S_2|<2$. In \sec{sec:Min} we found that $S^{\a\b}$ does not have a kinetic term around Minkowski space. So whether $S^{\a\b}$ leads to new degrees of freedom depends on the background, and a theory that contains it is unstable around some FLRW backgrounds. As $S^{\a\b}$ corresponds to the projectively invariant tensor $\dR_{(\a\b)}$, we again see that projective invariance does not guarantee stability.

As in Minkowski space, we do not obtain kinetic terms for $P^{\a\b}$. It thus remains an open question whether the corresponding tensor $\pR_{\a\b}$ can lead to new degrees of freedom, and if so, whether they are healthy.

\subsection{Zero torsion}

Let us now consider the case when the connection is taken to be symmetric a priori, \ie torsion vanishes, as often assumed in the Palatini formulation. This condition does not depend on the choice of the metric. Any projective transformation generates torsion (as \re{vectorpro} shows), so this assumption breaks projective invariance.\footnote{It would be possible to consider a symmetrised version of the projective transformation that maintains the zero torsion condition.} With zero torsion, $\accentset{\sim}{R}_{\a\b} = 2 R_{[\a\b]}$, so we can drop the dependence on $\accentset{\sim}{R}_{\a\b}$, setting $H^{\a\b}=0$.

When varying the action, we have to take into account that the connection is symmetric. From \re{variation} we see that this amounts to symmetrising $f_\a{}^{\b\c\d}$, and thus the equation of motion \re{gen_con_eom}, with respect to $\b$ and $\d$. To first order in the perturbations around Minkowski space, the equation of motion is
\bea \label{symmetric_eom_1st}
  && - \overset{q}{Q}_{\c\a\b} + \eta_{\c(\a} \overset{q}{\hat Q}_{\b)}+ \ha \eta_{\a\b} \overset{q}{Q}_{\c} - \ha\eta_{\c(\a} \overset{q}{Q}_{\b)} = 2\pd_\d\tf_{\c(\a\b)}{}^{\d} \ ,
\eea
with the solution
\bea \label{sym_Q_sol}
\!\!\!\!\!\!	\overset{q}{Q}_{\c\a\b} &=& -2\pd_\d\tf_{\c(\a\b)}{}^{\d} +\eta_{\a\b}\pd_\mu\tf_{\c\d}{}^{\d\mu}+\frac{2}{3}\pd_\mu\tf_{\d}{}_{(\a}{}^{\d\mu}\eta_{\b)\c}-\frac{1}{3}\eta_{\a\b}\pd_\mu\tf^{\d}{}_{\d\c}{}^{\mu}\el
	&&+\frac{2}{3}\pd_{\mu}\tf_{\d}{}^{\d}{}_{(\a}{}^{\mu}\eta_{\b)\c}-\frac{1}{3}\eta_{\a\b}\pd^\mu\tf^{\d}{}_{\c\d\mu} \el
	&=& \frac{1}{2}\pd_{(\a} S_{\b)\c} +\frac{1}{2}\pd_\c S_{\a\b} -\frac{1}{6}\eta_{\c(\a}\pd_{\b)}S^\mu{}_\mu -\frac{1}{6}\eta_{\a\b} S^\mu{}_\mu +\frac{1}{6}\eta_{\c(\a}\pd^\mu S_{\b)\mu} +\frac{1}{6}\eta_{\a\b} \pd^\mu S_{\c\mu} \el
	&&  + \frac{1}{2} \pd_{(\a} P_{\b)\c} + -\frac{1}{6}\eta_{\c(\a}\pd^\mu P_{\b)\mu} -\frac{1}{6}\eta_{\a\b}\pd^\mu P_{\c\mu} + \frac{2}{3}\eta_{\c(\a}\pd^\mu B_{\b)\mu} -\frac{1}{3}\eta_{\a\b}\pd^\mu B_{\c\mu} \  .
\eea
Substituting the solution \eqref{sym_Q_sol} into the action gives
\bea \label{action_zero_torsion}
S &=& \int \rmd^4 x \half \left[\vphantom{\half}\right. \lcR(q) +  \pd_\c \tf^{\a\b\c\d}\pd_\mu\tf_{\a(\b\d)}{}^{\mu} + \half \pd^\c\tf^{\a\b}{}_{\b\c}\pd_\mu\tf_{\a\d}{}^{\d\mu} - \half \pd_\c\tf^{\a\b\c\d}\pd_\mu\tf_{\b\a\d}{}^{\mu} \el
&&-\half\pd_\c\tf^{\a\b\c\d}\pd_\mu\tf_{\d\b\a}{}^\mu-\pd_\c\tf^{\a\b\c\d}\pd_\mu\tf_{\b\d\a}{}^{\mu}-\pd_\c\tf_{\a}{}^{\b\a\c}\pd_\mu\tf_{\b\d}{}^{\d\mu}+\pd_\b\tf_\a{}^{\a\b\c}\pd_\mu\tf_{\c\d}{}^{\d\mu}\el
&&+\frac{1}{6}\pd_\c\tf_\a{}^{\b\a\c}\pd^\mu\tf^\d{}_{\b\d\mu}-\frac{1}{3}\pd_\b\tf_\a{}^{\a\b\c}\pd^\mu\tf^\d{}_{\c\d\mu}-\frac{1}{6}\pd_\b\tf_\a{}^{\a\b\c}\pd^\mu\tf^\d{}_{\d\c\mu} \el
&& + \frac{\sg}{\sq} F -  (q^{\a\b}+B^{\a\b})\Sigma_{\a\b} -S^{\a\b}\Delta_{\a\b} -P^{\a\b}\Pi_{\a\b} \left.\vphantom{\half}\right] \ .
\eea
Finally, substituting $\tf_\a{}^{\b\c\d}$ from \eqref{def_gen_f_1st} with $H^{\a\b}=0$ into the action \eqref{action_zero_torsion}, performing an integration by parts, and diagonalising the $P_{\a\b}$ and $B_{\a\b}$ kinetic terms with the field redefinition $\hat B_{\a\b} = {B}_{\a\b} + 2 P_{\a\b}$, we get the action
\bea \label{action_2nd_ABdiag}
		S &=& \int \rmd^4 x \half \left[\vphantom{\half}\right. \lcR(q) + \frac{1}{6} \pd_\a \hat{B}^{\a\b}\pd_\c \hat{B}^{\c}_{\ \,\b} -\frac{1}{2} \pd_\a P^{\a\b}\pd_\c P^{\c}_{\ \,\b} - \frac{1}{16}\pd_\c P_{\a\b}\pd^\c P^{\a\b}+\frac{1}{16}\pd_\c S_{\a\b}\pd^\c S^{\a\b} \el
		&& -\frac{1}{48}\pd_\a S^\b{}_\b\pd^\a S^\c{}_\c -\frac{1}{12}\pd_\a S^{\a\b}\pd_\c S^\c{}_\b +\frac{1}{24}\pd_\b S^{\b\c}\pd_\c S^\a{}_\a - \frac{5}{4} \pd_\a P^{\a\b}\pd_\c S^\c{}_\b \el
		&& + \frac{2}{3}\pd_\a \hat{B}^{\a\b}\pd_\c S^\c{}_\b + \frac{\sg}{\sq} F -  (q^{\a\b}+\hat{B}^{\a\b}+2 P^{\a\b} )\Sigma_{\a\b} -S^{\a\b}\Delta_{\a\b} -P^{\a\b}\Pi_{\a\b} \left.\vphantom{\half}\right].
\eea
In contrast to the case when there was no a priori constraint on the connection, dependence on any of the Ricci-type tensors now leads to new degrees of freedom around Minkowski space.

Individually (\ie ignoring cross terms) the kinetic terms of $B_{\a\b}$ (corresponding to $R_{[\a\b]}$) are healthy, while the kinetic terms of $P_{\a\b}$ and $S_{\a\b}$ (corresponding to $\pR_{\a\b}$ and $\dR_{(\a\b)}$, respectively) lead to ghosts \cite{VANNIEUWENHUIZEN1973478}. In the case $P^{\a\b}=0$ ghost modes remain after diagonalising the $B^{\a\b}$ and $S^{\a\b}$ kinetic terms. In the case $P_{\a\b}=S_{\a\b}=0$, it is possible to solve the connection exactly, and the extra degrees of freedom in $B_{\a\b}$ correspond to one vector field with a positive kinetic term \cite{BeltranJimenez:2019acz}.
The theory can still be unstable if the field is tachyonic, which depends on the form of $F$.

\subsection{Zero non-metricity}

Let us now consider the case when non-metricity is taken to vanish a priori. This assumption is common in loop quantum gravity, and a theory that satisfies it is called Einstein--Cartan theory. Again, projective invariance is broken, and the constraint \re{procon} does not apply. Unlike the zero torsion condition, the assumption that non-metricity is zero depends on which metric we use: the original metric $g_{\a\b}$ or the new metric $q_{\a\b}$.

Let us first consider the case $\overset{g}{Q}_{\c\a\b}=0$. This is the most natural option, as the theory is originally defined with the metric $g_{\a\b}$. The corresponding disformation is zero, $J^\a{}_{\b\c}=0$, so $L^\a{}_\b{}_\c=K^\a{}_\b{}_\c$, where these quantities are defined with the decomposition \re{Con_decomp} in terms of $g_{\a\b}$. Writing the homothetic curvature tensor \re{Riccitensors} in terms of the Riemann tensor decomposition \re{Rie_dec} and using the property $K^{\a}{}_\b{}^{\c}=K^{[\a}{}_\b{}^{\c]}$ shows that $\hR_{\a\b}=0$. We thus have $H^{\a\b}=0$, just as in the zero torsion case. Furthermore, \re{ricci_av} shows that $\dR_{\a\b}=0$, so $S^{\a\b}=0$, and we can also set $P^{\a\b}=0$, as $\pR_{\a\b}$ is not independent. So only $B^{\a\b}$ remains.

What if we instead take the non-metricity defined with the new metric to be zero, $\overset{q}{Q}_{\c\a\b}=0$? We can now decompose the homothetic curvature tensor in terms of the new metric $q_{\a\b}$ and again find that $\hR_{\a\b}=0$, as discussed after \re{hRdec}. So we still have $H^{\a\b}=0$. In contrast, now $\dR_{\a\b}$ does not vanish in general. However, the only difference to the previous case is that the trace $\cR_{\a\b}$ of the Riemann tensor in \re{Riccitensors} is defined with $q^{\a\b}$ instead of $g^{\a\b}$. As we work to first order, only the background metric enters here, and if the background metrics are conformal to each other, we have to first order $\dR_{\a\b}=0$. In particular, this is true for perturbations around Minkowski space. So we can again set $S^{\a\b}=P^{\a\b}=0$, and only $B^{\a\b}$ remains.

Varying the action with respect to the connection and taking into account that the symmetric part is determined by the antisymmetric part when non-metricity is zero gives, according to \re{variation},
\bea \label{variation_torsion}
  \d S &=& \int \rmd^4 x \sq \left[ - \left( \nabla_\c + T_\c \right) \d^{\d}{}_\mu + \ha T^{\d}{}_{\c\mu} \right] f_\a{}^{\b\c\mu} \frac{\d\Gamma^\a_{\d\b}}{\d T^\sigma{}_{\rho\lambda}}\d T^\sigma{}_{\rho\lambda} \ .
\eea
Setting the variation to zero and expanding the equation of motion to first order in perturbations around Minkowski space, we get
\bea
  T_{[\b|\a|\c]} + 2\eta_{\a[\b} T_{\c]} + \frac{1}{2}T_{\a\b\c} &=&  \pd_\d\tf_{\a[\b\c]}{}^\d + \pd_\d\tf_{[\c|\a|\b]}{}^\d + \pd_\d\tf_{\c[\b\a]}{}^\d \ ,
\eea
with the solution
\bea
  T_{\c\a\b} &=& -2\pd_\d\tf_{[\a\b]\c}{}^\d + \eta_{\c[\a}\pd^\mu\tf^\d{}_{\b]\d\mu} - \eta_{\c[\a}\pd^\mu\tf_{\b]}{}^{\d}{}_{\d\mu} \ .
\eea
Inserting this back into the action and expanding to second order gives
\bea
	S &=& \int \rmd^4 x  \half \left[\vphantom{\half}\right. \lcR(q) 
  +\pd_\c\tf^{\a\b\c\d}\pd_\mu\tf_{\a(\b\d)}{}^\mu
  -\pd_\c\tf^{\a\b\c\d}\pd_\mu\tf_{\b(\a\d)}{}^\mu
  +\pd_\c\tf^{\a\b\c\d}\pd_\mu\tf_{[\d\b]\a}{}^\mu \el
  &&
  +\pd^\c\tf^{[\a\b]}{}_{\b\c}\pd_\mu\tf_{\a\d}{}^{\d\mu}
  +\pd^\c\tf^{\b\a}{}_{\b\c}\pd_\mu\tf_{[\d\a]}{}^{\d\mu}
  + \frac{\sg}{\sq} F -  (q^{\a\b}+B^{\a\b})\Sigma_{\a\b}\left.\vphantom{\half}\right] \ .
\eea
Substituting $\tf_\a{}^{\b\c\d}$ from \eqref{def_gen_f_1st} with $S^{\a\b}$, $P^{\a\b}$, and $H^{\a\b}$ set to zero, we get no kinetic terms for the new field $B^{\a\b}$. If the action is quadratic in $R_{[\a\b]}$, the full nonlinear theory has a ghost \cite{Hayashi:1980qp, Yo:2001sy, Vasilev:2017zyc, Barker:2023fem}. There is also a possible strong coupling problem due to the difference in dynamical degrees of freedom between the linearised case and the full theory.

\section{Conclusions} \label{sec:conc}

We have investigated the stability of non-degenerate actions that depend algebraically on Ricci-type tensors, \ie the first traces \eqref{Riccitensors} of the Riemann tensor. We performed a Legendre transformation and introduced auxiliary fields to make the action linear in the Riemann tensor. We then solved perturbatively for the connection and inserted it back into the action to shift the non-standard gravitational dynamics to the matter sector. We considered three cases: the general case when no constraints are imposed on the connection, the case with zero torsion, and the case with zero non-metricity. Our results are summarised in \tab{tab:dofs}.

\begin{table*}[t!]
  \begin{center}
  \begin{tabular}{|c|cc|c|c|}
  \hline
  Tensor & General case & & Zero torsion & Zero non-metricity \\
  & Minkowski & FLRW & Minkowski & Minkowski \\
  \hline
  $R_{(\a\b)}$ & no new dofs & - & no new dofs & no new dofs \\
  $R_{[\a\b]}$ & ghosts & - & healthy vector & no new dofs \\
  $\dR_{(\a\b)}$ & no new dofs & ghosts & ghosts & 0 \\
  $\pR_{\a\b}$ & no new dofs & no new dofs & ghosts & not independent \\
  $\hR_{\a\b}$ & \makecell{\\no new dofs, \\healthy vector,\\or ghosts} & - & not independent & 0 \\
  \hline
  \end{tabular}
  \end{center}
  \caption{Summary of which Ricci-type tensors lead to new degrees of freedom (dofs), and which of those are ghosts. An entry of 0 means that the tensor is zero, while ``not independent" means that it depends linearly on the other tensors. On the Minkowski background, the homothetic curvature tensor $\hR_{\a\b}$ can lead to a healthy vector, ghosts, or no new degrees of freedom, depending on the form of the function $F$. The new degrees of freedom due to $\dR_{(\a\b)}$ in the FLRW background in the general case can be stable or unstable, depending on the background. In the FLRW case we included $R_{(\a\b)}$ only via its trace and did not include $R_{[\a\b]}$.}
  \label{tab:dofs}
  \end{table*}

In the general case, dependence on $R_{[\a\b]}$, $\dR_{[\a\b]}$ or the projectively invariant combination $4R_{[\a\b]}-\hR_{\a\b}$ leads to ghosts around Minkowski space. Around Minkowski space neither of the tensors $\dR_{(\a\b)}$ nor $\pR_{\a\b}$ lead to new degrees of freedom. Dependence on $\hR_{\a\b}$ around Minkowski space can to a healthy vector field, ghosts, or no new degrees of freedom, depending on the form of the function $F$. Around FLRW space, $\pR_{\a\b}$ does not give new degrees of freedom, however $\dR_{(\a\b)}$ gives new degrees of freedom that are unstable for some backgrounds. Note that $\dR_{(\a\b)}$ is projectively invariant. This difference in the number of dynamical degrees of freedom for different backgrounds may point to a strong coupling problem around Minkowski space, casting doubt on the viability of a theory that depends on $\dR_{(\a\b)}$ \cite{Hindawi:1995cu, Golovnev:2018wbh, Pookkillath:2020iqq, BeltranJimenez:2020lee}. Our results in the general case are now consistent with \cite{Barker:2024ydb}. When we impose constraints on the connection, the results are quite different.

In the case with zero torsion, $\hR_{\a\b}$ is not an independent tensor. All other Ricci-type tensors apart from $R_{(\a\b)}$ now lead to new degrees of freedom. Due to couplings between the fields, the kinetic sector is complicated. Taken in isolation, $R_{[\a\b]}$ leads to a healthy vector field, as is well known, and true independent of the background \cite{Vitagliano:2010pq, Olmo:2013lta, BeltranJimenez:2019acz}. In isolation, the other Ricci-type tensors $\pR_{\a\b}$ and $\dR_{(\a\b)}$ lead to ghosts. If we drop $\pR_{\a\b}$, the conclusions about $R_{[\a\b]}$ and $\dR_{(\a\b)}$ hold. We did not diagonalise the kinetic sector when $\pR_{\a\b}$ is also present.

In the case with zero non-metricity we have $\hR_{\a\b}=0$. If non-metricity is defined with respect to the original metric $g_{\a\b}$, we also have $\dR_{\a\b}=0$. If it is defined with respect to the new metric $q_{\a\b}$, this holds to linear order around Minkowski space, but not in general. The only remaining Ricci-type tensor is $R_{\a\b}$. The symmetric part $R_{(\a\b)}$ does not lead to new degrees of freedom \cite{Afonso:2017, Galtsov:2018xuc, Annala:2021zdt, Olmo:2022rhf}, while $R_{[\a\b]}$ does not give new degrees of freedom around Minkowski space, but is known to have a ghost vector in the full theory \cite{Hayashi:1980qp, Yo:2001sy, Vasilev:2017zyc, Barker:2023fem}.

The only case with stable new degrees of freedom is when torsion is zero and the action depends only on the Ricci tensor, which was known already. As our focus is on whether a theory is healthy, we did not enumerate the number and type of all degrees of freedom, only whether some of them are unstable. We have in all cases required that the action is non-degenerate, \ie that the Legendre transformation is invertible. There can be degenerate actions that depend on some of the tensors excluded above but nevertheless give a stable theory.

The results demonstrate that in order to determine whether a theory can depend on some Ricci-type tensor, it is necessary to consider them in combination, as there can be cancellations. It remains open whether $\pR_{\a\b}$ ever leads to new degrees of freedom, and if so, whether they are healthy. From the fact that we do not get kinetic terms for it around Minkowski space or FLRW space we cannot conclude that this never happens. However, if there are new degrees of freedom around a different background, this may point to a strong coupling problem. Wrong sign kinetic terms from Ricci-type tensors can be cancelled by other terms, such as those that depend explicitly on the torsion \cite{Vasilev:2017zyc}. Our results show that projective invariance is not a sufficient condition for a theory to be ghost-free. The most general healthy Palatini theory remains to be determined.

\section*{Acknowledgments}

JA is supported by Academy of Finland projects 320123 and 345070. We thank Will Barker for pointing out errors in earlier versions of the paper. We also thank Will Barker, Tony Liimatainen, and Paolo Muratore-Ginannesch for useful discussions.

\appendix

\section{Solving the connection in the FLRW case} \label{app:con}

In this appendix we provide details on solving the connection in the FLRW background studied in section \refeq{sec:FLRW}. 

Substituting the decompositions \eqref{anz_bS} and \eqref{anz_bL} into the equations of motion for the connection at zeroth order gives four independent equations for the coefficients $L_i$. Due to projective symmetry the coefficient $L_2$ is left undetermined. One of the equations is simply $L_5=0$, corresponding to the fact that there is no parity-violating source. From the rest we solve for $L_1$, $L_3$, and $L_4$, obtaining
\bea
&&L_1 = -\frac{S_2}{4}\left[ 2H + \frac{S_2^\prime}{a(S_2+2)} \right] \ , \quad L_3 = -\frac{S_2}{4}\left[ 2H + \frac{S_2^\prime}{a(S_2-2)} \right] \el 
&&L_4 = -HS_2 - \frac{(S_2^2+4)S_2^\prime}{2a(S_2^2-4)} \ .
\eea
Substituting these back into \eqref{anz_bL}, we get the non-zero components of the background distortion:
\bea
  \bL^0{}_{00} &=& -\frac{2S_2^\prime}{S_2^2-4}-aL_2 \ , \qquad \qquad \quad \ \bL^i{}_{0j} = -a L_2\delta^i{}_j \el
\bL^i{}_{j0}&=&\left[ \frac{a^\prime}{2a}S_2 + \frac{S_2 S_2^\prime}{4(S_2-2)} \right]\delta^i{}_j \ , \quad \bL^0{}_{ij} = -\left[ \frac{a^\prime}{2a}S_2 + \frac{S_2 S_2^\prime}{4(S_2+2)} \right]\delta_{ij} \ .
\eea
Substituting this solution into the background action gives \eqref{bg_act}.

After finding the background, we need to solve the connection to first order, and substitute it back into the action to obtain the kinetic terms for the perturbations. The equation of motion for the connection expanded to first order \eqref{con1st} is in detail (indices are lowered with the background metric $\bq_{\a\b}$ and raised with its inverse)
\bea\label{con1st_full}
0&=& \Big( \bq^{\c\nu} \delta^\mu{}_\a \delta^\b{}_\sigma  + \bq^{\b\mu} \delta^\nu{}_\a \delta^\c{}_\sigma - \bq^{\mu\nu}\delta^{\c}{}_{\a}\delta^\b{}_\sigma  - \bq^{\b}{}^{\c}\delta^\mu{}_\sigma\delta^\nu{}_\a  + \tfrac{1}{2}\bq^{\b}{}^{\c}\bS_{\sigma}{}^{\mu}\delta^\nu{}_\a - \tfrac{1}{2}\bq^{\c\nu}\bS_{\a}{}^{\mu}\delta^\b{}_\sigma \el
&& + \tfrac{1}{2}\bq^{\mu\nu}\bS_{\a}{}^{\c}\delta^\b{}_\sigma - \tfrac{1}{2}\bq^{\b\mu}\bS^{\c}{}_{\sigma}\delta^\nu{}_\a - \tfrac{1}{2}\bS^{\mu\nu}\delta_{\a}{}^{\c}\delta^\b{}_\sigma  + \tfrac{1}{2}\bS^{\b}{}^{\mu}\delta^\c{}_\sigma\delta^\nu{}_\a  - \tfrac{1}{2}\bS^{\b}{}^{\c}\delta^\mu{}_\sigma \delta^\nu{}_\a \el 
&& + \tfrac{1}{2}\bS^{\c}{}^{\nu}\delta^\b{}_\sigma \delta^\mu{}_\a \Big) \dL^\sigma{}_{\mu\nu} + \rho_{\a}{}^{\b\c} \ ,
\eea
where the source term $\rho_{\a}{}^{\b\c}$ is
\bea\label{full_src}
\rho_{\a}{}^{\b\c} &=&  \frac{1}{2}\Big( \bL{}^{\b}{}^{\d}{}_{\d}\bS{}^{\c}{}^{\mu}\delta^\nu{}_{\a} - \bL{}^{\b}{}^{\d}{}^{\c}\bS{}_{\d}{}^{\mu}\delta^{\nu}{}_\a  + 2\bL{}^{\d}{}_{\d}{}_{\a}\bq^{\b\mu}\bq^{\c\nu}  - \bL{}^{\d}{}^{\lambda}{}_{\a}\bS{}_{\d}{}_{\lambda} \bq^{\b\mu}\bq^{\c\nu} - \bL{}^{\b}{}^{\d}{}^{\mu}\bS{}_{\a}{}^{\c}\delta^{\nu}{}_{\d} \el
&& - \bL{}^{\d}{}^{\b}{}_{\a}\bS{}^{\c}{}^{\mu}\delta^\nu{}_\d - 2\bL{}^{\c}{}^{\d}{}_{\a} \bq^{\b\mu}\delta^{\nu}{}_\d + \bL{}^{\d}{}^{\mu}{}_{\a}\bS{}^{\c}{}_{\d} \bq^{\b\nu}  - 2\bL{}^{\b}{}_{\a}{}^{\d} \bq^{\c\mu}\delta^{\nu}{}_{\d} + \bL{}^{\b}{}^{\d}{}^{\mu}\bS{}_{\a}{}_{\d}\bq^{\c\nu} \el 
&& + 2\bL{}^{\b}{}^{\d}{}^{\mu}\delta{}_{\a}{}^{\c} \delta^\nu{}_\d + \bL{}^{\d}{}^{\lambda}{}_{\a}\bS{}_{\lambda}{}^{\nu}\bq{}^{\b}{}^{\c} \delta^\mu{}_\d + \bq^{\b\mu}\bq^{\nu\c}\bcd{}_{\d}\bS{}_{\a}{}^{\d} - \delta^\mu{}_\a \bq{}^{\b}{}^{\c}  \bcd{}_{\lambda}\bS{}^{\nu}{}^{\lambda} \el
&& + \delta^\mu{}_\a \bcd{}^{\b}\bS{}^{\c\nu} - \bq^{\b\mu}\bq^{\d\nu}\bcd{}_{\d}\bS{}_{\a}{}^{\c} \Big) \dq_{\mu\nu} + \frac{1}{4}\Big( \bS^{\mu\nu}\bq{}^{\b}{}^{\c} \delta^\d{}_\a - \bS{}^{\b}{}^{\nu}\bq^{\c\d}\delta^\mu{}_\a   + \bS{}_{\a}{}^{\nu}\bq^{\mu\c} \bq^{\b\d} \el
&& + \bS^{\mu\nu}\bq^{\b\d}\delta{}_{\a}{}^{\c} - \bS{}_{\a}{}^{\nu}\bq^{\b\mu} \bq^{\d\c} - \bS{}^{\b}{}^{\d}\bq^{\mu\nu}\delta{}_{\a}{}^{\c} - \bS{}_{\a}{}^{\d}\bq{}^{\b}{}^{\c}\bq^{\mu\nu} + 2\bS{}^{\c}{}^{\d}\bq^{\b\nu}\delta^\mu{}_{\a} \el 
  && + \bS{}^{\b}{}^{\d} \bq^{\nu\c}\delta^\mu{}_\a + \bS{}_{\a}{}^{\d} \bq^{\mu\b}\bq^{\nu\c} - 2\bS{}^{\nu}{}^{\d}\bq{}^{\b}{}^{\c}\delta^\mu{}_\a - 2\bS{}^{\nu}{}^{\d}\bq^{\b\mu}\delta{}_{\a}{}^{\c} +\bS{}^{\b}{}^{\nu} \bq^{\mu\c} \delta^\d{}_\a \Big) \bcd{}_{\d} \dq_{\mu\nu} \el
&&+\frac{1}{2}\Big(  \bL{}^{\b}{}^{\d}{}_{\d} \bq_{\a\mu}\delta^\c{}_\nu - \bL{}^{\b}{}_{\nu}{}^{\c}\bq_{\a\mu}  - \bL{}^{\d}{}_{\d}{}_{\a}\delta^\b{}_\mu\delta^\c{}_\nu + \bL{}^{\c}{}_{\nu}{}_{\a}\delta^\b{}_\mu + \bL{}^{\b}{}_{\a}{}_{\nu}\delta^\c{}_\mu  - \bL{}_{\nu}{}^{\b}{}_{\a}\delta^\c{}_\mu \el
&& + \bL{}_{\nu}{}_{\mu}{}_{\a}\bq{}^{\b}{}^{\c} - \bL{}^{\b}{}_{\nu}{}_{\mu}\delta{}_{\a}{}^{\c} \Big) \dS^{\mu\nu} + \frac{1}{2}\Big(\delta^\d{}_\a\delta^{\b}{}_\mu \delta^\c{}_\nu  + \bq_{\a\mu}\bq^{\d\b}\delta^\nu{}_\c  - \bq_{\a\mu}\bq{}^{\b}{}^{\c} \delta^\d{}_\nu \el
  && - \delta^{\c}{}_{\a} \delta^\b{}_\mu\delta^\d{}_\nu  \Big) \bcd_\d \dS^{\mu\nu} + \frac{1}{2}\Big( \bL{}^{\b}{}_{\nu}{}^{\c}\bq_{\a\mu} - \bL{}^{\b}{}^{\d}{}_{\d} \bq_{\a\mu}\delta^\c{}_\nu - \bL{}^{\d}{}_{\d}{}_{\a}\delta^\b{}_\mu \delta^\c{}_\nu + \bL{}^{\c}{}_{\nu}{}_{\a}\delta^\b{}_\mu \el
&& - \bL{}^{\b}{}_{\a}{}_{\nu}\delta^\c{}_\mu - \bL{}_{\nu}{}^{\b}{}_{\a}\delta^\c{}_\mu  - \bL{}_{\mu}{}_{\nu}{}_{\a}\bq{}^{\b}{}^{\c} + \bL{}^{\b}{}_{\mu}{}_{\nu}\delta{}_{\a}{}^{\c}\Big)\dP^{\mu\nu} + \frac{1}{2}\Big(\delta^\d{}_\a \delta^\b{}_\mu\delta^\c{}_\nu  \el
&& - \bq_{\a\mu}\bq^{\d\b}\delta^\c{}_\nu  + \bq_{\a\mu}\bq{}^{\b}{}^{\c} \delta^\d{}_\nu - \delta{}_{\a}{}^{\c}\delta^\b{}_\mu \delta^\d{}_\nu  \Big) \bcd_\d \dP^{\mu\nu} \ .
\eea
Due to projective invariance the equation of motion leaves one vector undetermined. We choose this to be the torsion vector $T_\a=L^\c{}_{\a\c}-L^{\c}{}_{\c\a}$. Note that $\rho_{\a}{}^{\a\b} = 0$, so $\rho{}_{i}{}^{i}{}^{0}{} = -\rho_0{}^{00}$ and $\rho{}_{j}{}^{j}{}^{i} = -\rho{}_{0}{}^{0}{}^{i}$. The equations of motion \eqref{con1st_full} then read component by component
\bea \label{Lback}
  0 &=& -\tfrac{1}{2}(S_2 - 2)\dL{}^{i}{}_{i}{}_{0} + \tfrac{1}{2}( S_2 + 2 )\dL{}_{0}{}^{i}{}_{i} - \rho_0{}^{0}{}_{0} \el
  0 &=& \tfrac{1}{2}(S_2 - 2)\dL{}^{i}{}_{00} - \tfrac{1}{2}( S_2 + 2 )\dL{}_{0}{}_{0}{}^{i} - \rho{}_{0}{}_{0}{}^{i} \el
  0 &=& -\tfrac{1}{2}a^2( S_2 + 2 )\dL{}^{i}{}^{k}{}_{k} - \tfrac{1}{2}( S_2 + 2 )\dL{}_{0}{}^{i}{}_{0} - \rho{}_{0}{}^{i}{}_{0} \el
  0 &=& \dL{}^{i}{}_{j}{}_{0} -\dL{}^{k}{}_{k}{}_{0}\delta{}^{i}{}_{j} - \tfrac{1}{2}(S_2+2)\dL^0{}_{00}\delta{}^{i}{}_{j} + \tfrac{1}{2}( S_2 + 2 )\dL{}^{i}{}_{0}{}_{j} + \rho{}_{0}{}^{i}{}_{j} \el
  0 &=& -\tfrac{1}{2}(S_2 - 2)\dL{}^{k}{}_{k}{}_{i} + \tfrac{1}{2}(S_2 - 2)\dL{}^{0}{}_{i}{}_{0} - \rho{}_{i}{}^{0}{}_{0} \el
  0 &=&\dL{}^{0}{}_{i}{}_{j} - \dL{}^{0}{}^{k}{}_{k}\bq_{ij} - \tfrac{1}{2}(S_2 - 2)\dL{}_{j}{}^{0}{}_{i} + \tfrac{1}{2}(S_2-2)\dL^{00}{}_{0}\bq_{ij} + \rho{}_{i}{}^{0}{}{}_{j} \el
  0 &=& \tfrac{1}{2}(S_2 - 2)\dL{}^{j}{}_{i}{}_{0} - \tfrac{1}{2}( S_2 + 2 )\dL{}_{0}{}^{j}{}_{i} - \rho{}_{i}{}^{j}{}_{0}{} \el
  0 &=& \dL{}^{j}{}_{i}{}_{k} + \dL{}_{k}{}^{j}{}_{i} - \dL{}^{l}{}_{l}{}_{i}\delta{}^{j}{}_{k}  - \tfrac{1}{2}(S_2+2)\dL{}^{0}{}_{0}{}_{i}\delta{}^{j}{}_{k} - \tfrac{1}{2}(S_2-2)\dL{}^{j}{}_{00}\delta{}_{i}{}_{k} - \dL{}^{j}{}^{l}{}_{l}\bq{}_{i}{}_{k}  + \rho{}_{i}{}^{j}{}_{k} \ . \el
\eea
It is straightforward to solve the components of the connection, with the result
\bea
  \label{Lpert1} \dL^0{}_{00} &=& \frac{1}{12( S_2 + 2 )}[( S_2 + 6 )\rho{}_{0}{}^{i}{}_{i} +( S_2 + 2 )\rho{}_{i}{}_{0}{}{}^{i} - 4\rho_0{}^{0}{}_{0} ] + \frac{1}{3}T_0 \\
  \dL{}^{i}{}_{00} &=& -\frac{1}{2(S_2-2)^2(S_2 + 2)}\Big[  a^2(S_2^2 -4)\rho{}^{i}{}^{k}{}_{k} - a^2(S_2^2 - 4)\rho{}_{j}{}^{i}{}^{j} - 2( S_2 + 2 )\rho{}^{i}{}_{00} \el
  && - 2(S_2 - 2)\rho{}_{0}{}^{i}{}_{0} - 2(S_2^2 - 4)\rho{}_{0}{}_{0}{}^{i} \Big] \\
  \dL{}^{0}{}_{i}{}_{0} &=& \frac{1}{6(S_2^2-4)( S_2 + 2 )}\Big[(S_2^2-4) ( S_2 + 3 ) \rho{}_{i}{}^{k}{}_{k} + ( S_2 + 1 )(S_2^2-4)\rho{}_{k}{}_{i}{}^{k} \el
  && +2(S_2+2)(S_2+3)\rho{}_{i0}{}^{0} -2(S_2-2)(3S_2+5)\rho{}_{0}{}_{i}{}^{0}  +2(S_2^2-4)(2S_2+3)\rho{}_{0}{}^{0}{}_{i} \Big] \el
  && + \frac{1}{3}T_i \\
\dL{}^{0}{}_{0}{}_{i} &=& \frac{1}{2(S_2^2-4)( S_2 + 2 )}\Big[(S_2^2 - 4)\rho{}_{i}{}^{k}{}_{k} - (S_2^2 - 4)\rho{}_{k}{}_{i}{}^{k} + 2( S_2 + 2 )\rho{}_{i0}{}^{0} \el
  && + 2(S_2^2-4)\rho{}_{0}{}^{0}{}_{i} - 2( S_2 -2)\rho{}_{0}{}_{i}{}^{0} \Big] \\
  \dL{}^{i}{}_{j}{}_{0} &=& \frac{1}{4( S_2 - 2)}\Big[ ( S_2 + 2 )\delta{}^{i}{}_{j}\rho{}_{k}{}_{0}{}{}^{k} -2( S_2 + 2 )\rho{}^{i}{}_{0}{}{}_{j} - 2(S_2-2)\rho{}_{0}{}_{j}{}^{i} +( S_2 - 2)\delta{}^{i}{}_{j}\rho{}_{0}{}^{k}{}_{k} \el
  && +4\rho{}_{j}{}^{i}{}_{0}{} \Big] \\
  \dL{}^{j}{}_{0}{}_{i} &=& \frac{1}{12(S_2^2 - 4)}\Big[  12( S_2 + 2 )\rho{}_{i}{}_{0}{}{}^{j} -24\rho{}^{j}{}_{i}{}_{0}{} + (S_2^2-4)\delta^{j}{}_{i}\rho{}_{k}{}_{0}{}{}^{k} - 12( S_2 - 2 )\rho{}_{0}{}^{j}{}_{i} \el
&& +(S_2-2)(S_2+6)\delta^{j}{}_{i}\rho{}_{0}{}^{k}{}_{k} + 4( S_2 + 4 )\delta^{j}{}_{i}\rho_0{}^{00}   \Big] + \frac{1}{3}\delta^{j}{}_{i}T_{0} \\
\dL{}^{0}{}_{i}{}_{j} &=& \frac{1}{4( S_2 + 2 )}\Big[ (S_2+2)\bq{}_{i}{}_{j}\rho{}_{k}{}^{0}{}{}^{k} - 2( S_2 + 2 )\rho{}_{i}{}^{0}{}{}_{j} -2(S_2 - 2)\rho{}^{0}{}_{j}{}_{i} - 4 \rho{}_{j}{}_{i}{}^{0} \el
  && + ( S_2 - 2 ) \bq{}_{i}{}_{j}\rho{}^{0}{}_{k}{}^{k} \Big]
\eea
\bea
  \label{Lpert8} \dL{}^{i}{}_{j}{}_{k} &=& \frac{1}{12(S_2^2-4)( S_2 + 2 )}\Big[3( S_2 + 2 )\bq{}_{j}{}_{k}\Big\{ ( S_2^2 - 4 ) ( - \rho{}^{i}{}^{l}{}_{l} + \rho{}_{l}{}^{i}{}^{l}) - 2( S_2 + 2 )\rho{}^{i}{}_{00} \el 
  && - 2(S_2 - 2)\rho{}_{0}{}^{i}{}_{0} \Big\}  - 3 ( S_2 + 2 )  \Big\{  2 ( S_2^2 - 4 ) ( - \rho{}^{i}{}_{k}{}_{j} + \rho{}_{j}{}^{i}{}_{k} + \rho{}_{k}{}_{j}{}^{i}) \el
  && + \delta^{i}{}_{j}  \big[ ( S_2^2 - 4 ) (   \rho{}_{l}{}_{k}{}^{l} - \rho{}_{k}{}^{l}{}_{l} ) + 2 ( S_2 + 2 ) \rho{}_{k}{}^{0}{}_{0} + 2(S_2 - 2)\rho{}_{0}{}_{k}{}^{0}  \big]  \Big\} \el
  && + 2\delta^{i}{}_{k}  \Big\{  (S_2^2-4)(S_2+3)\rho{}_{j}{}^{l}{}_{l} + (S_2^2-4)( S_2 + 1 )\rho{}_{l}{}_{j}{}^{l} - 2(S_2+2)(2S_2+3)\rho{}_{j}{}^{0}{}_{0}\el
  && + 2(S_2 - 2)\rho{}_{0}{}_{j}{}^{0} + S_2 ( S_2^2 - 4 ) \rho{}_{0}{}^{0}{}_{j} \Big\} \Big] + \frac{1}{3}\delta^{i}{}_{k}T_j \ .
\eea
Because of projective invariance we can choose the projective mode freely. We choose it so that the torsion vector vanishes, $T_\a=0$. For the background decomposition \eqref{anz_bL} this translates into $L_2 = L_3$. Substituting \re{Lback} and \re{Lpert1}--\re{Lpert8}, with the source \eqref{full_src}, into the action \eqref{action_FLRW_2} gives the kinetic terms.

\section{Pseudovector and vector degrees of freedom} \label{app:general}

\subsection{General action}

In the case when the connection is unconstrained and $H^{\a\b}$ and possibly $B^{\a\b}$ (corresponding to $\hR_{\a\b}$ and $R_{[\a\b]}$, respectively) are included, the stability of the theory can depend on the form of the function $F$. Expanding $F$ to second order in the fields around Minkowski space and solving $\Sigma_{\a\b}$, $\Delta_{\a\b}$, $\Pi_{\a\b}$, and $\Theta_{\a\b}$ from \eqref{redefs}--\eqref{redefH}, we have
\bea \label{gen_act_expanded_F}
  S &=& \int \rmd^4 x \half \Bigg[ \lcR  - \frac{7}{2}\pd_\a {B}^{\a\b}\pd^\c {B}_{\c\b} -\frac{1}{4}\pd_\c {B}_{\a\b}\pd^\c {B}^{\a\b}  - \frac{166}{3}\pd_\a {H}^{\a\b}\pd^\c {H}_{\c\b} -\frac{82}{3} \pd_\a {B}^{\a\b}\pd^\c {H}_{\c\b} \el
  && + \frac{1}{3} ( - 2 T_\b + \pd^\c S_{\c\b} + \pd^\c P_{\c\b} ) ( \pd_\a B^{\a\b} + 4\pd_\a H^{\a\b} ) + b_2 B^{\a\b}B_{\a\b} + s_1 S^{\a}{}_\a S^{\b}{}_\b + s_2 S^{\a\b} S_{\a\b} \el
  && + p_2 P^{\a\b}P_{\a\b} + h_2 H^{\a\b}H_{\a\b} + b_h B^{\a\b}H_{\a\b} + b_p B^{\a\b}P_{\a\b} + h_p H^{\a\b}P_{\a\b}\Bigg] \ .
\eea
The constants are
\begin{align}
  b_2 &= \frac{1}{4 A}(4\pi_2 \theta_2 - \theta_\pi^2) \ , \quad b_h = \frac{1}{2 A}(\sigma_\pi \theta_\pi - 2 \sigma_\theta) \ , \el
  h_2 &= -\frac{1}{4A}(\sigma_\pi^2 - 4\sigma_2 \pi_2) \ , \quad h_p = \frac{1}{2 A}(\sigma_\pi \sigma_\theta - 2 \sigma_2 \theta_\pi) \ , \el
  p_2 &= -\frac{1}{4A}(\sigma_\theta^2 - 4 \sigma_2 \theta_2) \ , \quad b_p = \frac{1}{2A}(\sigma_\theta \theta_\pi - 2\sigma_\pi \theta_2) \ , \el
  s_1 &= \frac{\delta_r^2 r_2}{(\delta_r^2 - 4 \delta_2 r_3)^2} \ , \quad s_2 = \frac{r_3}{\delta_r^2 - 4 \delta_2 r_3} \ ,
\end{align}
where $A \equiv \sigma_\theta^2 \pi_2 + \sigma_\pi^2 \theta_2 - \sigma_\pi \sigma_\theta \theta_\pi + \sigma_2(\theta_\pi^2 - 4\pi_2\theta_2)$, and $r_2$, $r_3$, $\sigma_2$, $\delta_2$, $\pi_2$, $\theta_2$, $\theta_\pi$, $\sigma_\pi$, $\sigma_\theta$, and $\delta_r$ are constants coming from the expansion of $F$ to second order:
\bea
  F &=& r_1 \Sigma^\a{}_\a + r_2 (\Sigma^\a{}_\a)^2 + r_3 \Sigma_{(\a\b)}\Sigma^{(\a\b)} + \sigma_2 \Sigma_{[\a\b]}\Sigma^{[\a\b]} + \pi_2 \Pi_{\a\b}\Pi^{\a\b} + \theta_2 \Theta_{\a\b}\Theta^{\a\b} \el 
  && + \delta_2 \Delta_{\a\b}\Delta^{\a\b} + \sigma_\pi \Sigma_{[\a\b]}\Pi^{\a\b} + \sigma_\theta \Sigma_{[\a\b]}\Theta^{\a\b} + \theta_\pi \Theta_{\a\b}\Pi^{\a\b} + \delta_r \Delta_{\a\b}\Sigma^{(\a\b)}.
\eea
In the case when there is no $B^{\a\b}$ field (corresponding to $\Sigma_{[\a\b]}$) in the action, we have $b_2=b_h=b_p=0$, and $h_2$, $p_2$, and $h_p$ are
\begin{align}
  h_2 = \frac{\pi_2}{\theta_\pi^2 - 4\pi_2\theta_2} \ , \quad p_2 = \frac{\theta_2}{\theta_\pi^2 - 4\pi_2\theta_2} \ , \quad h_p = -\frac{\theta_\pi}{\theta_\pi^2 - 4\pi_2\theta_2} \ .
\end{align}
The coefficients $s_1$ and $s_2$ are unchanged.

\subsection{The case with $B^{\a\b}$}

Let us first consider the case when $B^{\a\b}$ appears in the action. It is enough for us to show that at least one new degree of freedom is unstable. It is simplest to establish that there is a ghost in the pseudovector sector of the action. This way we do not have to consider the Lagrange multiplier type term (related to the projective mode), which constrains the vector sector, complicating the analysis. We decompose the action into the different spin-parity sectors using the set of projectors for a general rank 2 tensor $\{ \mathcal{P}^{2^+},$ $ \mathcal{P}^{0^+}_s,$ $ \mathcal{P}^{0^+}_\omega,$ $ \mathcal{P}^{1^-}_m,$ $ \mathcal{P}^0_{s\omega},$ $ \mathcal{P}^0_{\omega s},$ $ \mathcal{P}^{1^+}_b,$ $ \mathcal{P}^{1^-}_e,$ $ \mathcal{P}^{1^-}_{me},$ $ \mathcal{P}^{1^-}_{em} \}$, where spin $J$ and parity $P$ are denoted as $J^P$; for the details of the projectors, see \eg the appendix of \cite{VANNIEUWENHUIZEN1973478} or appendix B of \cite{Buoninfante:2016iuf}. The pseudovector sector ($J^P = 1^+$) is completely decoupled from the other sectors because the action is parity-invariant. Noting that $k^\a (\mathcal{P}^{1^+}_{b})_{\a\b\c\d} = 0$, where $k^\a$ is the momentum and $(\mathcal{P}^{1^+}_{b})_{\a\b\c\d}$ is the $1^+$ spin projector, we see that the only kinetic term in \eqref{gen_act_expanded_F} that contributes to this sector is $\pd_\c B_{\a\b}\pd^\c B^{\a\b}$. The terms $b_h B^{\a\b}H_{\a\b}$ and $b_p B^{\a\b}P_{\a\b}$ can only contribute to the mass term. In the case when all cross terms are absent, the spin $1^+$ sector of the field equations with a source term $B_{\a\b}J_B^{\a\b}$ in momentum space gives
\begin{align}
  \big(b_2 - \tfrac{1}{4} k^2\big) (\mathcal{P}^{1^+}_{b})_{\a\b\c\d} B^{\c\d} = (\mathcal{P}^{1^+}_{b})_{\a\b\c\d} J^{\c\d} \ .
\end{align}
From this we can read off that the propagator is $1/(b_2 - \tfrac{1}{4} k^2)$. The residue of the pole of this propagator ($\mathrm{Res}_{k^2=4b_2}[ 1/(b_2 - \tfrac{1}{4} k^2) ] = -4$) is negative, so the spin $1^+$ sector contains a ghost. The above analysis would be exactly the same if the action depended on $\dR_{[\a\b]}$ instead of $R_{[\a\b]}$, since the resulting relevant kinetic term that comes from $\dR_{[\a\b]}$ has the same form.

The details are more complicated once we include the cross terms. However, the essential point remains that the stability of the pseudovector is still determined solely by the sign of the kinetic term $\pd_\c {B}_{\a\b}\pd^\c {B}^{\a\b}$. With the cross terms the field equations can be written in matrix form as
\begin{align}\label{pseudo_vec_mix}
  \begin{bmatrix}
    b_2 -\tfrac{1}{4}k^2 & \tfrac{1}{2} b_h & \tfrac{1}{2} b_p \\
    \tfrac{1}{2} b_h     & h_2              & \tfrac{1}{2} h_p \\
    \tfrac{1}{2} b_p     & \tfrac{1}{2} h_p & p_2
  \end{bmatrix}
  \begin{bmatrix}
    B^{\c\d} \\
    H^{\c\d} \\
    P^{\c\d}
  \end{bmatrix}
  (\mathcal{P}^{1^+}_{b})_{\a\b\c\d}
  =
  \begin{bmatrix}
    J_B^{\c\d} \\
    J_H^{\c\d} \\
    J_P^{\c\d}
  \end{bmatrix}
  (\mathcal{P}^{1^+}_{b})_{\a\b\c\d} \ ,
\end{align}
where $J_H^{\c\d}$ and $J_P^{\c\d}$ are the sources for $H^{\a\b}$ and $P^{\a\b}$, respectively. The propagator for the pseudovector sector is obtained by inverting the kinetic matrix. We denote the kinetic matrix in \eqref{pseudo_vec_mix} by $M$ and its inverse by $M^{-1}$. The poles of the propagator can be found from the zeroes of the determinant of $M$,
\begin{align}
  \det M = \tfrac{1}{4}\big[ b_h b_p h_p -b_p^2 h_2 -b_h^2 p_2 - (h_p^2-4h_2 p_2)(b_2 -\tfrac{1}{4}k^2) \big] = 0 \ .
\end{align}
This gives the pole $k^2 = -4( b_h b_p h_p -b_p^2 h_2 -h_p^2 b_2 -b_h^2 p_2 + 4b_2 h_2 p_2 )/(h_p^2- 4h_2 p_2)\equiv m^2$. Now the residue of the propagator at the pole is
\begin{align}
  \underset{k^2 = m^2}{\mathrm{Res}}\mathrm{Tr}[M^{-1}] = -4\Bigg[ 1 + \frac{ ( b_h h_p - 2 b_p h_2 )^2 + ( b_p h_p - 2 b_h p_2 )^2}{  (h_p^2 - 4h_2 p_2)^2 } \Bigg] \ ,
\end{align}
where the overall factor $-4$ comes from the factor $-\tfrac{1}{4}$ in front of the kinetic term $\pd_\c {B}_{\a\b}\pd^\c {B}^{\a\b}$. The term inside the square brackets is positive, so the residue of the pole of the propagator is negative, and $B^{\a\b}$ gives a pseudovector ghost, whether or not the action depends also on other fields.

\subsection{The case without $B^{\a\b}$}

If the $B^{\a\b}$ field is absent, the stability of the theory depends on the form of $F$. The equations of motion for the fields are
\begin{align}
  0 &=\frac{332}{3}\pd_\a\pd^{[\c}H^{\b]\a} + 2h_2 H^{\c\b} + h_p P^{\c\b} + \frac{4}{3}(\pd_\a\pd^{[\c}P^{\b]\a}+\pd_\a\pd^{[\c}S^{\b]\a}) - \frac{8}{3}\pd^{[\c}T^{\b]} \ , \el
  0 &= s_1 S^\a{}_\a \eta^{\c\b} + s_2 S^{\c\b} + \frac{2}{3}\pd_\a\pd^{(\c}H^{\b)\a} \ , \el
  0 &= h_p H^{\c\b} + 2p_2 P^{\c\b} + \frac{4}{3}\pd_\a\pd^{[\c}H^{\b]\a} \ , \el
  0 &= \pd^\a H_{\a\b} \ .
\end{align}
Let us first consider the case when $4s_1+s_2\neq0$, $h_p^2- 4 h_2 p_2\neq0$, $p_2\neq0$. The equations of motion then simplify to 
\begin{align} \label{noB_simplified_eoms}
  S^{\c\b}=0 \ , \ P^{\c\b} = -\frac{h_p}{2p_2}H^{\c\b} \ , \ H^{\c\b} = \frac{8}{3} \Big( 2h_2 - \tfrac{h_p^2}{2p_2}\Big)^{-1} \pd^{[\c}T^{\b]} \ , \  \pd^\b\pd_{[\c}T_{\b]} = 0 \ .
\end{align}
The action thus reduces to
\begin{align} \label{noB_vec_action}
  S &= \int \rmd^4 x \half \Bigg[ \lcR + \frac{16}{9} \Big(2h_2-\tfrac{h_p^2}{2p_2}\Big)^{-1} \pd^{[\a}T^{\b]}\pd_{[\a}T_{\b]} \Bigg] \ .
\end{align}
We see that there is a new vector field that is stable if $\tfrac{h_p^2}{4p_2} > h_2$ and a ghost if $\tfrac{h_p^2}{4p_2} < h_2$. So the stability of theory depends on the form of $F$. Let us now consider the special cases when some of the above combinations of constants vanish.

If $s_2 = -4 s_1$, then only the traceless part of $S^{\a\b}$ appears in the action, so the equations of motion yield $S^{\a\b} = \tfrac{1}{4}\eta^{\a\b} S^\c{}_\c$, and the trace $S^\c{}_\c$ remains unconstrained. Nevertheless, the term $\pd_\a \pd^{[\c}S^{\b]\a} = \tfrac{1}{4} \pd^{[\b}\pd^{\c]}S^\a{}_\a = 0$ vanishes and the rest of the equations in \eqref{noB_simplified_eoms} remain the same. So the conclusions do not change.

If $\tfrac{h_p^2}{4p_2}=h_2$, we have instead $\pd^{[\c}T^{\b]}=0$, $P^{\c\b} = -\tfrac{h_p}{2p_2} H^{\c\b}$ and $\pd^\b H_{\a\b} = 0$, and there are no new degrees of freedom.

If $p_2=0$, $h_p\neq0$, then from the third and fourth equation in \eqref{noB_simplified_eoms} we have $H^{\c\b}=0$, so the new terms in the action vanish, and there are no new degrees of freedom.

If $p_2=h_p=0$, the situation is slightly more involved. In this case the equations of motion simplify to $\pd_\a H^{\a\b} = 0$ and $H^{\a\b} = \tfrac{1}{2h_2}( \tfrac{8}{3}\pd^{[\c}T^{\b]} - \tfrac{4}{3}\pd_\a\pd^{[\c}P^{\b]\a})$. With the change of variables $\hat{P^\a} \equiv \pd_\b P^{\b\a} + 2T^\a$ the action can then be written as
\begin{align} \label{p2_hp_zero_action}
  S &= \int \rmd^4 x \half \Bigg( \lcR + \frac{4}{9h_2} \pd^{[\a}\hat{P}^{\b]}\pd_{[\a}\hat{P}_{\b]} \Bigg) \ .
\end{align}
Thus there is again a propagating vector field, which can be stable or a ghost depending on the form of $F$.

So, in the case with no $B^{\a\b}$ field, we can have either no new degrees of freedom, one healthy vector, or a ghost vector, depending on the form of $F$.

\bibliographystyle{JHEP}
\bibliography{ricci}

\providecommand{\href}[2]{#2}\begingroup\raggedright\begin{thebibliography}{10}

\bibitem{Birrell:1982ix}
N.~D. Birrell and P.~C.~W. Davies, \emph{{Quantum Fields in Curved Space}}.
\newblock Cambridge Monographs on Mathematical Physics. Cambridge Univ. Press,
  Cambridge, UK, 1984,
  \href{http://dx.doi.org/10.1017/CBO9780511622632}{10.1017/CBO9780511622632}.

\bibitem{Zwiebach:1985uq}
B.~Zwiebach, \emph{{Curvature Squared Terms and String Theories}},
  \href{http://dx.doi.org/10.1016/0370-2693(85)91616-8}{\emph{Phys. Lett. B}
  {\bfseries 156} (1985) 315--317}.

\bibitem{Zumino:1985dp}
B.~Zumino, \emph{{Gravity Theories in More Than Four-Dimensions}},
  \href{http://dx.doi.org/10.1016/0370-1573(86)90076-1}{\emph{Phys. Rept.}
  {\bfseries 137} (1986) 109}.

\bibitem{Deser:1986xr}
S.~Deser and A.~N. Redlich, \emph{{String Induced Gravity and Ghost Freedom}},
  \href{http://dx.doi.org/10.1016/0370-2693(86)90177-2}{\emph{Phys. Lett. B}
  {\bfseries 176} (1986) 350}.

\bibitem{Eichhorn:2020mte}
A.~Eichhorn, \emph{{Asymptotically safe gravity}},  in \emph{{57th
  International School of Subnuclear Physics}: {In Search for the Unexpected}},
  2, 2020, \href{https://arxiv.org/abs/2003.00044}{{\ttfamily 2003.00044}}.

\bibitem{Stelle:1977ry}
K.~S. Stelle, \emph{{Classical Gravity with Higher Derivatives}},
  \href{http://dx.doi.org/10.1007/BF00760427}{\emph{Gen. Rel. Grav.} {\bfseries
  9} (1978) 353--371}.

\bibitem{Simon:1990ic}
J.~Z. Simon, \emph{{Higher Derivative Lagrangians, Nonlocality, Problems and
  Solutions}}, \href{http://dx.doi.org/10.1103/PhysRevD.41.3720}{\emph{Phys.
  Rev.} {\bfseries D41} (1990) 3720}.

\bibitem{MuellerHoissen:1991}
F.~Mueller-Hoissen, \emph{{Higher derivative versus second order field
  equations}}, {\emph{Annalen Phys.} {\bfseries 48} (1991) 543--557}.

\bibitem{Nunez:2004ts}
A.~Nunez and S.~Solganik, \emph{{Ghost constraints on modified gravity}},
  \href{http://dx.doi.org/10.1016/j.physletb.2005.01.015}{\emph{Phys. Lett. B}
  {\bfseries 608} (2005) 189--193},
  [\href{https://arxiv.org/abs/hep-th/0411102}{{\ttfamily hep-th/0411102}}].

\bibitem{Chiba:2005nz}
T.~Chiba, \emph{{Generalized gravity and ghost}},
  \href{http://dx.doi.org/10.1088/1475-7516/2005/03/008}{\emph{JCAP} {\bfseries
  03} (2005) 008}, [\href{https://arxiv.org/abs/gr-qc/0502070}{{\ttfamily
  gr-qc/0502070}}].

\bibitem{Woodard:2006nt}
R.~P. Woodard, \emph{{Avoiding dark energy with 1/r modifications of gravity}},
  \href{http://dx.doi.org/10.1007/978-3-540-71013-4_14}{\emph{Lect. Notes
  Phys.} {\bfseries 720} (2007) 403--433},
  [\href{https://arxiv.org/abs/astro-ph/0601672}{{\ttfamily
  astro-ph/0601672}}].

\bibitem{Donoghue:2021eto}
J.~F. Donoghue and G.~Menezes, \emph{{Ostrogradsky instability can be overcome
  by quantum physics}},
  \href{http://dx.doi.org/10.1103/PhysRevD.104.045010}{\emph{Phys. Rev. D}
  {\bfseries 104} (2021) 045010},
  [\href{https://arxiv.org/abs/2105.00898}{{\ttfamily 2105.00898}}].

\bibitem{Sotiriou:2008rp}
T.~P. Sotiriou and V.~Faraoni, \emph{{f(R) Theories Of Gravity}},
  \href{http://dx.doi.org/10.1103/RevModPhys.82.451}{\emph{Rev. Mod. Phys.}
  {\bfseries 82} (2010) 451--497},
  [\href{https://arxiv.org/abs/0805.1726}{{\ttfamily 0805.1726}}].

\bibitem{Kobayashi:2019hrl}
T.~Kobayashi, \emph{{Horndeski theory and beyond: a review}},
  \href{http://dx.doi.org/10.1088/1361-6633/ab2429}{\emph{Rept. Prog. Phys.}
  {\bfseries 82} (2019) 086901},
  [\href{https://arxiv.org/abs/1901.07183}{{\ttfamily 1901.07183}}].

\bibitem{Einstein:1925}
A.~Einstein, \emph{Einheitliche feldtheorie von gravitation und
  elektrizit{\"a}t}, {\emph{Verlag der Koeniglich-Preussichen Akademie der
  Wissenschaften} {\bfseries 22} (07, 1925) 414--419}.

\bibitem{Hehl:1976}
F.~W. Hehl, P.~{Von Der Heyde}, G.~D. Kerlick and J.~M. Nester, \emph{{General
  Relativity with Spin and Torsion: Foundations and Prospects}},
  \href{http://dx.doi.org/10.1103/RevModPhys.48.393}{\emph{Rev. Mod. Phys.}
  {\bfseries 48} (1976) 393--416}.

\bibitem{Hehl:1978}
F.~W. {Hehl} and G.~D. {Kerlick}, \emph{{Metric-affine variational principles
  in general relativity. I - Riemannian space-time}},
  \href{http://dx.doi.org/10.1007/BF00760141}{\emph{General Relativity and
  Gravitation} {\bfseries 9} (Aug., 1978) 691--710}.

\bibitem{Hehl:1981}
F.~W. {Hehl}, E.~A. {Lord} and L.~L. {Smalley}, \emph{{Metric-affine
  variational principles in general relativity II. Relaxation of the Riemannian
  constraint}}, \href{http://dx.doi.org/10.1007/BF00756364}{\emph{General
  Relativity and Gravitation} {\bfseries 13} (Nov., 1981) 1037--1056}.

\bibitem{Randono:2005}
A.~Randono, \emph{{A Note on parity violation and the Immirzi parameter}},
  \href{https://arxiv.org/abs/hep-th/0510001}{{\ttfamily hep-th/0510001}}.

\bibitem{Freidel:2005sn}
L.~Freidel, D.~Minic and T.~Takeuchi, \emph{{Quantum gravity, torsion, parity
  violation and all that}},
  \href{http://dx.doi.org/10.1103/PhysRevD.72.104002}{\emph{Phys. Rev.}
  {\bfseries D72} (2005) 104002},
  [\href{https://arxiv.org/abs/hep-th/0507253}{{\ttfamily hep-th/0507253}}].

\bibitem{Perez:2005pm}
A.~Perez and C.~Rovelli, \emph{{Physical effects of the Immirzi parameter}},
  \href{http://dx.doi.org/10.1103/PhysRevD.73.044013}{\emph{Phys. Rev.}
  {\bfseries D73} (2006) 044013},
  [\href{https://arxiv.org/abs/gr-qc/0505081}{{\ttfamily gr-qc/0505081}}].

\bibitem{Mercuri:2006um}
S.~Mercuri, \emph{{Fermions in Ashtekar-Barbero connections formalism for
  arbitrary values of the Immirzi parameter}},
  \href{http://dx.doi.org/10.1103/PhysRevD.73.084016}{\emph{Phys. Rev.}
  {\bfseries D73} (2006) 084016},
  [\href{https://arxiv.org/abs/gr-qc/0601013}{{\ttfamily gr-qc/0601013}}].

\bibitem{Mercuri:2006wb}
S.~Mercuri, \emph{{Nieh-Yan Invariant and Fermions in Ashtekar-Barbero-Immirzi
  Formalism}},  in \emph{{Recent developments in theoretical and experimental
  general relativity, gravitation and relativistic field theories. Proceedings,
  11th Marcel Grossmann Meeting, MG11, Berlin, Germany, July 23-29, 2006. Pt.
  A-C}}, pp.~2794--2796, 2006,
  \href{https://arxiv.org/abs/gr-qc/0610026}{{\ttfamily gr-qc/0610026}}.

\bibitem{Bojowald:2007nu}
M.~Bojowald and R.~Das, \emph{{Canonical gravity with fermions}},
  \href{http://dx.doi.org/10.1103/PhysRevD.78.064009}{\emph{Phys. Rev.}
  {\bfseries D78} (2008) 064009},
  [\href{https://arxiv.org/abs/0710.5722}{{\ttfamily 0710.5722}}].

\bibitem{Kazmierczak:2008}
M.~Kazmierczak, \emph{{Einstein-Cartan gravity with Holst term and fermions}},
  \href{http://dx.doi.org/10.1103/PhysRevD.79.064029}{\emph{Phys. Rev.}
  {\bfseries D79} (2009) 064029},
  [\href{https://arxiv.org/abs/0812.1298}{{\ttfamily 0812.1298}}].

\bibitem{Shaposhnikov:2020gts}
M.~Shaposhnikov, A.~Shkerin, I.~Timiryasov and S.~Zell, \emph{{Higgs inflation
  in Einstein-Cartan gravity}},
  \href{http://dx.doi.org/10.1088/1475-7516/2021/10/E01}{\emph{JCAP} {\bfseries
  02} (2021) 008}, [\href{https://arxiv.org/abs/2007.14978}{{\ttfamily
  2007.14978}}].

\bibitem{Shaposhnikov:2020aen}
M.~Shaposhnikov, A.~Shkerin, I.~Timiryasov and S.~Zell, \emph{{Einstein-Cartan
  Portal to Dark Matter}},
  \href{http://dx.doi.org/10.1103/PhysRevLett.127.169901}{\emph{Phys. Rev.
  Lett.} {\bfseries 126} (2021) 161301},
  [\href{https://arxiv.org/abs/2008.11686}{{\ttfamily 2008.11686}}].

\bibitem{Vasilev:2017zyc}
T.~B. Vasilev, J.~A.~R. Cembranos, J.~G. Valcarcel and P.~Mart\'\i{}n-Moruno,
  \emph{{Stability in quadratic torsion theories}},
  \href{http://dx.doi.org/10.1140/epjc/s10052-017-5331-6}{\emph{Eur. Phys. J.
  C} {\bfseries 77} (2017) 755},
  [\href{https://arxiv.org/abs/1706.07080}{{\ttfamily 1706.07080}}].

\bibitem{BeltranJimenez:2019acz}
J.~Beltrán~Jiménez and A.~Delhom, \emph{{Ghosts in metric-affine higher order
  curvature gravity}},
  \href{http://dx.doi.org/10.1140/epjc/s10052-019-7149-x}{\emph{Eur. Phys. J.
  C} {\bfseries 79} (2019) 656},
  [\href{https://arxiv.org/abs/1901.08988}{{\ttfamily 1901.08988}}].

\bibitem{BeltranJimenez:2020sqf}
J.~Beltr\'an~Jim\'enez and A.~Delhom, \emph{{Instabilities in metric-affine
  theories of gravity with higher order curvature terms}},
  \href{http://dx.doi.org/10.1140/epjc/s10052-020-8143-z}{\emph{Eur. Phys. J.
  C} {\bfseries 80} (2020) 585},
  [\href{https://arxiv.org/abs/2004.11357}{{\ttfamily 2004.11357}}].

\bibitem{Percacci:2020ddy}
R.~Percacci and E.~Sezgin, \emph{{New class of ghost- and tachyon-free metric
  affine gravities}},
  \href{http://dx.doi.org/10.1103/PhysRevD.101.084040}{\emph{Phys. Rev. D}
  {\bfseries 101} (2020) 084040},
  [\href{https://arxiv.org/abs/1912.01023}{{\ttfamily 1912.01023}}].

\bibitem{Jimenez-Cano:2022sds}
A.~Jim\'enez-Cano and F.~J. Maldonado~Torralba, \emph{{Vector stability in
  quadratic metric-affine theories}},
  \href{http://dx.doi.org/10.1088/1475-7516/2022/09/044}{\emph{JCAP} {\bfseries
  09} (2022) 044}, [\href{https://arxiv.org/abs/2205.05674}{{\ttfamily
  2205.05674}}].

\bibitem{Donoghue:2019ecz}
J.~F. Donoghue and G.~Menezes, \emph{{Arrow of Causality and Quantum Gravity}},
  \href{http://dx.doi.org/10.1103/PhysRevLett.123.171601}{\emph{Phys. Rev.
  Lett.} {\bfseries 123} (2019) 171601},
  [\href{https://arxiv.org/abs/1908.04170}{{\ttfamily 1908.04170}}].

\bibitem{Donoghue:2020mdd}
J.~F. Donoghue and G.~Menezes, \emph{{Quantum causality and the arrows of time
  and thermodynamics}},
  \href{http://dx.doi.org/10.1016/j.ppnp.2020.103812}{\emph{Prog. Part. Nucl.
  Phys.} {\bfseries 115} (2020) 103812},
  [\href{https://arxiv.org/abs/2003.09047}{{\ttfamily 2003.09047}}].

\bibitem{Afonso:2017}
V.~I. Afonso, C.~Bejarano, J.~{Beltran Jimenez}, G.~J. Olmo and E.~Orazi,
  \emph{{The trivial role of torsion in projective invariant theories of
  gravity with non-minimally coupled matter fields}},
  \href{http://dx.doi.org/10.1088/1361-6382/aa9151}{\emph{Class. Quant. Grav.}
  {\bfseries 34} (2017) 235003},
  [\href{https://arxiv.org/abs/1705.03806}{{\ttfamily 1705.03806}}].

\bibitem{Galtsov:2018xuc}
D.~Gal'tsov and S.~Zhidkova, \emph{{Ghost-free Palatini derivative
  scalar\textendash{}tensor theory: Desingularization and the speed test}},
  \href{http://dx.doi.org/10.1016/j.physletb.2019.01.061}{\emph{Phys. Lett. B}
  {\bfseries 790} (2019) 453--457},
  [\href{https://arxiv.org/abs/1808.00492}{{\ttfamily 1808.00492}}].

\bibitem{Annala:2021zdt}
J.~Annala and S.~R{\"a}s{\"a}nen, \emph{{Inflation with R
  (\ensuremath{\alpha}\ensuremath{\beta}) terms in the Palatini formulation}},
  \href{http://dx.doi.org/10.1088/1475-7516/2021/09/032}{\emph{JCAP} {\bfseries
  09} (2021) 032}, [\href{https://arxiv.org/abs/2106.12422}{{\ttfamily
  2106.12422}}].

\bibitem{Olmo:2022rhf}
G.~J. Olmo and D.~Rubiera-Garcia, \emph{{Some recent results on Ricci-based
  gravity theories}},
  \href{http://dx.doi.org/10.1142/S0218271822400120}{\emph{Int. J. Mod. Phys.
  D} {\bfseries 31} (2022) 2240012},
  [\href{https://arxiv.org/abs/2203.04116}{{\ttfamily 2203.04116}}].

\bibitem{Vitagliano:2010pq}
V.~Vitagliano, T.~P. Sotiriou and S.~Liberati, \emph{{The dynamics of
  generalized Palatini Theories of Gravity}},
  \href{http://dx.doi.org/10.1103/PhysRevD.82.084007}{\emph{Phys. Rev. D}
  {\bfseries 82} (2010) 084007},
  [\href{https://arxiv.org/abs/1007.3937}{{\ttfamily 1007.3937}}].

\bibitem{Olmo:2013lta}
G.~J. Olmo and D.~Rubiera-Garcia, \emph{{Importance of torsion and invariant
  volumes in Palatini theories of gravity}},
  \href{http://dx.doi.org/10.1103/PhysRevD.88.084030}{\emph{Phys. Rev. D}
  {\bfseries 88} (2013) 084030},
  [\href{https://arxiv.org/abs/1306.4210}{{\ttfamily 1306.4210}}].

\bibitem{Alvarez:2017spt}
E.~Alvarez, J.~Anero and S.~Gonzalez-Martin, \emph{{Quadratic gravity in first
  order formalism}},
  \href{http://dx.doi.org/10.1088/1475-7516/2017/10/008}{\emph{JCAP} {\bfseries
  10} (2017) 008}, [\href{https://arxiv.org/abs/1703.07993}{{\ttfamily
  1703.07993}}].

\bibitem{Alvarez:2018lrg}
E.~Alvarez, J.~Anero, S.~Gonzalez-Martin and R.~Santos-Garcia, \emph{{Physical
  content of Quadratic Gravity}},
  \href{http://dx.doi.org/10.1140/epjc/s10052-018-6250-x}{\emph{Eur. Phys. J.
  C} {\bfseries 78} (2018) 794},
  [\href{https://arxiv.org/abs/1802.05922}{{\ttfamily 1802.05922}}].

\bibitem{Marzo:2021esg}
C.~Marzo, \emph{{Ghost and tachyon free propagation up to spin 3 in Lorentz
  invariant field theories}},
  \href{http://dx.doi.org/10.1103/PhysRevD.105.065017}{\emph{Phys. Rev. D}
  {\bfseries 105} (2022) 065017},
  [\href{https://arxiv.org/abs/2108.11982}{{\ttfamily 2108.11982}}].

\bibitem{Aoki:2019rvi}
K.~Aoki and K.~Shimada, \emph{{Scalar-metric-affine theories: Can we get
  ghost-free theories from symmetry?}},
  \href{http://dx.doi.org/10.1103/PhysRevD.100.044037}{\emph{Phys. Rev.}
  {\bfseries D100} (2019) 044037},
  [\href{https://arxiv.org/abs/1904.10175}{{\ttfamily 1904.10175}}].

\bibitem{Rasanen:2018b}
S.~R{\"a}s{\"a}nen, \emph{{Higgs inflation in the Palatini formulation with
  kinetic terms for the metric}},
  \href{http://dx.doi.org/10.21105/astro.1811.09514}{\emph{Open J. Astrophys.}
  {\bfseries 2} (2019) 1}, [\href{https://arxiv.org/abs/1811.09514}{{\ttfamily
  1811.09514}}].

\bibitem{Delhom:2021bvq}
A.~Delhom, \emph{{Theoretical and Observational Aspecs in Metric-Affine
  Gravity: A field theoretic perspective}}, Ph.D. thesis, Valencia U., 2021.
\newblock \href{https://arxiv.org/abs/2201.09789}{{\ttfamily 2201.09789}}.

\bibitem{Belenchia:2016bvb}
A.~Belenchia, M.~Letizia, S.~Liberati and E.~D. Casola, \emph{{Higher-order
  theories of gravity: diagnosis, extraction and reformulation via non-metric
  extra degrees of freedom\textemdash{}a review}},
  \href{http://dx.doi.org/10.1088/1361-6633/aaa4ab}{\emph{Rept. Prog. Phys.}
  {\bfseries 81} (2018) 036001},
  [\href{https://arxiv.org/abs/1612.07749}{{\ttfamily 1612.07749}}].

\bibitem{Hindawi:1995cu}
A.~Hindawi, B.~A. Ovrut and D.~Waldram, \emph{{Nontrivial vacua in higher
  derivative gravitation}},
  \href{http://dx.doi.org/10.1103/PhysRevD.53.5597}{\emph{Phys. Rev. D}
  {\bfseries 53} (1996) 5597--5608},
  [\href{https://arxiv.org/abs/hep-th/9509147}{{\ttfamily hep-th/9509147}}].

\bibitem{Golovnev:2018wbh}
A.~Golovnev and T.~Koivisto, \emph{{Cosmological perturbations in modified
  teleparallel gravity models}},
  \href{http://dx.doi.org/10.1088/1475-7516/2018/11/012}{\emph{JCAP} {\bfseries
  11} (2018) 012}, [\href{https://arxiv.org/abs/1808.05565}{{\ttfamily
  1808.05565}}].

\bibitem{Pookkillath:2020iqq}
M.~C. Pookkillath, A.~De~Felice and A.~A. Starobinsky, \emph{{Anisotropic
  instability in a higher order gravity theory}},
  \href{http://dx.doi.org/10.1088/1475-7516/2020/07/041}{\emph{JCAP} {\bfseries
  07} (2020) 041}, [\href{https://arxiv.org/abs/2004.03912}{{\ttfamily
  2004.03912}}].

\bibitem{BeltranJimenez:2020lee}
J.~Beltr\'an~Jim\'enez and A.~Jim\'enez-Cano, \emph{{On the strong coupling of
  Einsteinian Cubic Gravity and its generalisations}},
  \href{http://dx.doi.org/10.1088/1475-7516/2021/01/069}{\emph{JCAP} {\bfseries
  01} (2021) 069}, [\href{https://arxiv.org/abs/2009.08197}{{\ttfamily
  2009.08197}}].

\bibitem{Iosifidis:2018zjj}
D.~Iosifidis, A.~C. Petkou and C.~G. Tsagas, \emph{{Torsion/non-metricity
  duality in f(R) gravity}},
  \href{http://dx.doi.org/10.1007/s10714-019-2539-9}{\emph{Gen. Rel. Grav.}
  {\bfseries 51} (2019) 66},
  [\href{https://arxiv.org/abs/1810.06602}{{\ttfamily 1810.06602}}].

\bibitem{Magnano:1987zz}
G.~Magnano, M.~Ferraris and M.~Francaviglia, \emph{{Nonlinear gravitational
  Lagrangians}}, \href{http://dx.doi.org/10.1007/BF00760651}{\emph{Gen. Rel.
  Grav.} {\bfseries 19} (1987) 465}.

\bibitem{Ferraris:1988zz}
M.~Ferraris, M.~Francaviglia and G.~Magnano, \emph{{Do non-linear metric
  theories of gravitation really exist?}},
  \href{http://dx.doi.org/10.1088/0264-9381/5/6/002}{\emph{Class. Quant. Grav.}
  {\bfseries 5} (1988) L95}.

\bibitem{Magnano:1990qu}
G.~Magnano, M.~Ferraris and M.~Francaviglia, \emph{{Legendre transformation and
  dynamical structure of higher derivative gravity}},
  \href{http://dx.doi.org/10.1088/0264-9381/7/4/007}{\emph{Class. Quant. Grav.}
  {\bfseries 7} (1990) 557--570}.

\bibitem{Koga:1998un}
J.-i. Koga and K.-i. Maeda, \emph{{Equivalence of black hole thermodynamics
  between a generalized theory of gravity and the Einstein theory}},
  \href{http://dx.doi.org/10.1103/PhysRevD.58.064020}{\emph{Phys. Rev. D}
  {\bfseries 58} (1998) 064020},
  [\href{https://arxiv.org/abs/gr-qc/9803086}{{\ttfamily gr-qc/9803086}}].

\bibitem{PhysRevD.56.7769}
Y.~N. Obukhov, E.~J. Vlachynsky, W.~Esser and F.~W. Hehl, \emph{Effective
  einstein theory from metric-affine gravity models via irreducible
  decompositions},
  \href{http://dx.doi.org/10.1103/PhysRevD.56.7769}{\emph{Phys. Rev. D}
  {\bfseries 56} (Dec, 1997) 7769--7778}.

\bibitem{VANNIEUWENHUIZEN1973478}
P.~{van Nieuwenhuizen}, \emph{On ghost-free tensor lagrangians and linearized
  gravitation},
  \href{http://dx.doi.org/https://doi.org/10.1016/0550-3213(73)90194-6}{\emph{Nuclear
  Physics B} {\bfseries 60} (1973) 478 -- 492}.

\bibitem{Hayashi:1980qp}
K.~Hayashi and T.~Shirafuji, \emph{{Gravity From Poincare Gauge Theory of the
  Fundamental Particles. 4. Mass and Energy of Particle Spectrum}},
  \href{http://dx.doi.org/10.1143/PTP.64.2222}{\emph{Prog. Theor. Phys.}
  {\bfseries 64} (1980) 2222}.

\bibitem{Yo:2001sy}
H.-J. Yo and J.~M. Nester, \emph{{Hamiltonian analysis of Poincare gauge
  theory: Higher spin modes}},
  \href{http://dx.doi.org/10.1142/S0218271802001998}{\emph{Int. J. Mod. Phys.
  D} {\bfseries 11} (2002) 747--780},
  [\href{https://arxiv.org/abs/gr-qc/0112030}{{\ttfamily gr-qc/0112030}}].

\bibitem{Barker:2023fem}
W.~Barker and S.~Zell, \emph{{Einstein-Proca theory from the Einstein-Cartan
  formulation}},
  \href{http://dx.doi.org/10.1103/PhysRevD.109.024007}{\emph{Phys. Rev. D}
  {\bfseries 109} (2024) 024007},
  [\href{https://arxiv.org/abs/2306.14953}{{\ttfamily 2306.14953}}].

\bibitem{Barker:2024ydb}
W.~Barker and C.~Marzo, \emph{{Particle spectra of general Ricci-type Palatini
  or metric-affine theories}},
  \href{http://dx.doi.org/10.1103/PhysRevD.109.104017}{\emph{Phys. Rev. D}
  {\bfseries 109} (2024) 104017},
  [\href{https://arxiv.org/abs/2402.07641}{{\ttfamily 2402.07641}}].

\bibitem{Buoninfante:2016iuf}
L.~Buoninfante, \emph{{Ghost and singularity free theories of gravity}},
  \href{https://arxiv.org/abs/1610.08744}{{\ttfamily 1610.08744}}.

\end{thebibliography}\endgroup

\end{document}